\documentclass[12pt, draftclsnofoot, onecolumn]{IEEEtran}
\usepackage{amsmath,graphicx}
\usepackage{amsfonts}
\usepackage{enumerate}
\usepackage{bm}
\usepackage{algorithm} 
\usepackage{algorithmic} 
\usepackage{mathrsfs}
\usepackage{stfloats}
\usepackage{enumerate}
\usepackage{epstopdf}
\usepackage{subfigure}
\usepackage{float}
\usepackage{amsbsy}
\usepackage{amsmath}
\usepackage{amssymb}
\usepackage{color}
\usepackage{ntheorem}
\usepackage{cite}
\theorembodyfont{\upshape}
\theoremheaderfont{\rmfamily\itshape}
\theoremseparator{:}
\theoremstyle{remark}

\newtheorem{theo}{\hspace{1em}Theorem}
\newtheorem{lemma}{\hspace{1em}Corollary}
\newtheorem{case}{\hspace{1em}Lemma}
\newtheorem*{pproof}{\hspace{2em}Proof}

\begin{document}
\IEEEoverridecommandlockouts
\title{\huge{Modular Extremely Large-Scale
Array Communication: Near-Field Modelling and
Performance Analysis}\vspace{-1pt}}
\author{\IEEEauthorblockN{Xinrui~Li, Haiquan~Lu, Yong Zeng,  \emph{Member, IEEE}, Shi Jin, \emph{Senior Member, IEEE}, and Rui Zhang, \emph{Fellow, IEEE}
\thanks{This work was supported by the National Key R$\&$D Program of China with Grant number 2019YFB1803400, the National Natural Science Foundation of China under Grant numbers 62071114 and 61041104, the Fundamental Research Funds for the Central Universities of China under grant numbers 3204002004A2 and 2242022k30005.}
\thanks{X. Li, H. Lu, Y. Zeng, and S. Jin are with the National Mobile Communications Research
Laboratory and Frontiers Science Center for Mobile Information Communication and Security, Southeast University, Nanjing 210096, China. Y. Zeng is also
with the Purple Mountain Laboratories, Nanjing 211111, China (e-mail: \{230218659, 230208088, yong\_zeng, jinshi\}@seu.edu.cn). (\emph{Corresponding author: Yong Zeng}.)}
\thanks{R. Zhang is with the Chinese University of Hong Kong, Shenzhen, and Shenzhen Research Institute of Big Data, Shenzhen, China 518172.  He is also with the Department of Electrical and Computer Engineering, National University of Singapore, Singapore 117583 (e-mail: rzhang@cuhk.edu.cn).}
}}
\maketitle
\vspace{-1cm}

\begin{abstract}
  This paper investigates wireless communications based on a new antenna array architecture,
  termed \textit{modular extremely large-scale array (XL-array)},
  where an extremely large number of antenna elements are regularly arranged on a common platform in a modular manner. Each module consists of a flexible/moderate number of antenna elements,
  and different modules are separated with an inter-module spacing that is typically much larger than the inter-element spacing/signal wavelength for ease of deployment.
  By properly modelling the variations of signal phase, amplitude and projected aperture across different array modules/elements, we develop the new channel model and analyze the signal-to-noise ratio (SNR) performance of the modular XL-array based communications. Under the practical non-uniform spherical wave (NUSW) model, the closed-form expression of the maximum achievable SNR is derived in terms of key geometric parameters,
  including the total planar array size, module separation distances along each dimension, as well as the user's location in the three-dimensional (3D) space.
  Besides, the asymptotic SNR scaling laws are revealed as the number of modules along different dimensions goes to infinity. Moreover, we show that our developed near-field modelling and performance analysis include the existing ones for the collocated XL-array, the far-field uniform plane wave (UPW) model, as well as the one-dimensional (1D) modular extremely large-scale uniform linear array (XL-ULA) as special cases.
  Extensive simulation results are provided to validate our obtained results.
\end{abstract}

\section{Introduction}

As a key enabling technology for spectral-efficient communications, massive multiple-input multiple-output (MIMO) has been realized in the fifth-generation (5G) wireless communication networks \cite{Andrews2014, Zhang2020, Bjornson2019}.
To support the ambitious goals of 6G, such as ultra-high
throughput, ultra-low latency, and ultra-high reliability \cite{Bii2019, Rappaport2019, Tong2021}, several promising transmission technologies have been studied, including extremely large-scale MIMO (XL-MIMO) \cite{Lu2021, Zeng2021, Wang20211, Amiri2018, Bjornson2019}, Terahertz communication \cite{Akyildiz2016, Wang2021},
 large intelligent surface (LIS) \cite{Hu2021,Dardari2021}, and intelligent reflecting surface (IRS)/reconfigurable intelligent surface (RIS) \cite{Wu2021, Ozdogan2020, Pei2021}.
In particular, XL-MIMO that dramatically boosts antenna numbers and physical size beyond the current massive MIMO systems is expected to be able to significantly increase the system spectral efficiency and spatial resolution, which has attracted fast-growing attention recently. Besides XL-MIMO, other terms have also been used in the literature to refer to this technology, such as extremely large aperture arrays (ELAAs) \cite{Bjornson2019}, extremely
large aperture massive MIMO (xMaMIMO) \cite{Amiri2018}, and ultra-massive MIMO (UM-MIMO) \cite{Akyildiz2016}.\par

As the number of antennas keeps increasing, there are in general two MIMO architectures to accommodate the physically and electrically large antenna array, namely \textit{collocated extremely large-scale array (XL-array)} \cite{Lu2021, Zeng2022} and \textit{distributed antenna system (DAS)} \cite{Choi2020, Bjornson2019}. For collocated XL-array architecture,
the antenna elements are arranged regularly on a common platform with adjacent elements typically separated by the wavelength scale,
like standard antenna arrays.
When the antenna size becomes large and/or the link distance is small, the conventional uniform plane wave (UPW) based channel model is no longer valid for XL-array communications \cite{Zeng2022,Hu2021}.
Instead, the more general non-uniform spherical wave (NUSW) characteristics should be taken into account to more accurately model the variations of signal amplitude and phase across array elements. Some preliminary efforts along this direction have been made in \cite{Magoarou2020, Botond2020, Lu2021, Wei2021, Dar2020, Bj2020, Zeng2021}.
For example, in \cite{Lu2021}, based on the NUSW model, a collocated extremely large-scale uniform linear array (XL-ULA) was investigated for the single-user near-field communication.
In \cite{Dar2020} and \cite{Wei2021}, the effect of distance and phase variations across array elements has been
revealed for XL-MIMO channel estimation.
By taking into consideration the variations of signal amplitude, projected aperture, and the polarization mismatch, the authors in \cite{Bj2020} studied the near-field channel modelling for XL-MIMO communication.
Furthermore, in \cite{Zeng2021}, the authors pursued a generic three dimensional (3D) channel modelling that takes into account both elevation and azimuth angle of arrival/departure (AoA/AoD) for collocated extremely large-scale uniform planar array (XL-UPA) communication.
However, since the conventional collocated XL-array architecture in \cite{Lu2021,Wei2021,Dar2020,Bj2020,Zeng2021, Wang2021} requires
a contiguous platform for deployment, the array size is typically limited by the available space of the mounting structure in practice \cite{Bjornson2019}.
On the other hand, for DAS, the antennas are distributed over a large geographical area with multiple separated sites,
which are inter-connected by the backhaul/fronthaul links to perform joint signal processing and provide uniformly good performance ubiquitously \cite{Choi2020}.
Some representative examples
of DAS architecture include network MIMO \cite{kara2006}, cloud radio
access network (C-RAN) \cite{Yoon2015} and virtual MIMO \cite{Chang2016}. Recently, a novel DAS architecture termed \textit{cell-free massive MIMO} was proposed, which comprises a large number of distributed access points (APs) to simultaneously serve the users over a wide area \cite{Nayebi2015, Ngo2017, Ngo20172}.
However, DAS usually requires not only a large number of sites for antenna deployment,
but also the high backhaul/fronthaul capacity
and sophisticated coordination among different sites \cite{Ngo2017,Ngo20172}.\par

\begin{figure}[t]
\begin{centering}
\includegraphics[scale=0.3]{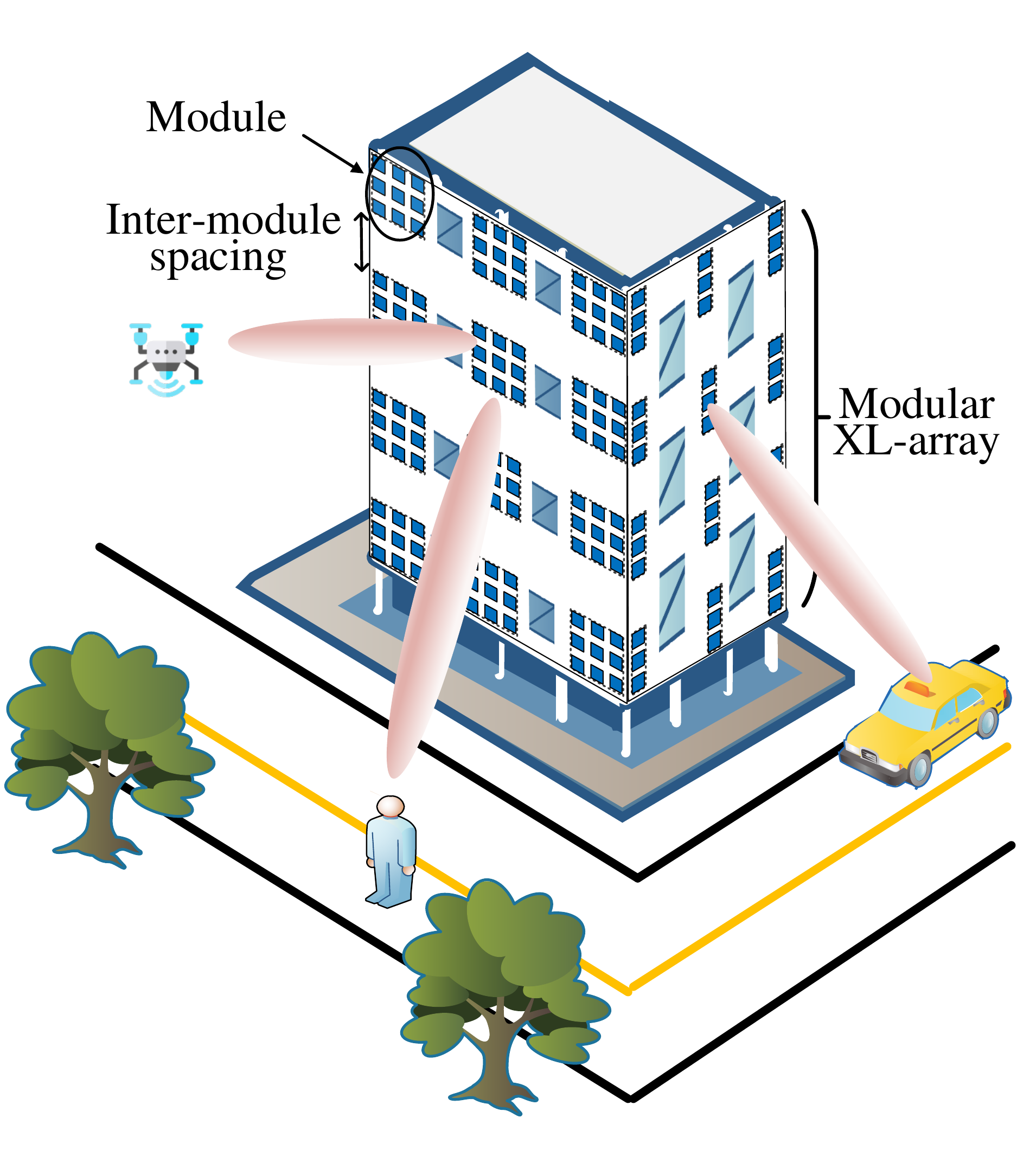}
\vspace{-0.3cm}
\caption{An illustration of modular XL-arrays mounted on building facades.} \label{picture11}
\end{centering}
\vspace{-0.5cm}
\end{figure}

To complement the existing collocated XL-array and DAS architectures,
in this paper, we investigate a novel modular array architecture to commodate extremely large arrays,
termed \textit{modular XL-array}. As illustrated in Fig. \ref{picture11}, for modular XL-array,
the array elements are regularly deployed on a common platform in a modular manner, like
Lego-type building blocks \cite{Jeon2021, Li2022}. Each module consists of a flexible/moderate number of array elements,
and different modules are separated by an inter-module spacing that is typically much larger than the inter-element spacing/signal wavelength, so as to enable conformal capability with the actual mounting structure. For example, modular XL-array can be embedded into facades of office buildings, airports, shopping malls, etc.,
where different modules are separated by windows. Note that the modular design can also be directly extended to IRS/RIS, where high passive beamforming gain usually requires extremely large-scale surface aperture \cite{JJ2021}. Compared to the collocated XL-array architecture, modular XL-array is more flexible for deployment. Furthermore, it may also help to improve the communication coverage, especially for cell-edge users \cite{Jeon2021}. Note that to support the large-scale deployment of modular XL-array, existing cost-effective array techniques can be employed, such as analog beamforming, hybrid analog/digital beamforming, and low-resolution analog-to-digital converters (ADCs), etc.
On the other hand, compared to DAS architecture, modular XL-array may achieve joint signal processing without having to perform sophisticated inter-site signal exchange or coordination, which may ease the requirement of synchronization and reduce hardware cost associated with the backhaul/fronthaul links for DAS \cite{Li2022}. Furthermore, as different modules of modular XL-array share a common site, it may ease the complex network planning and site selection.
However, it is worth mentioning that the aforementioned three array architectures may fit for different application scenarios. For example, collocated and modular XL-array may be used for supporting cellular hot spot, while DAS is a good candidate for providing uniformly good services everywhere over a relatively large area \cite{Bjornson2019, Li2022}. Therefore, the three array architectures are expected to be complementary to each other, and their choices are dependent on the practical scenarios.\par
The modular concept for antenna arrays was first introduced in the antenna design community \cite{Ashkenazy1982, Mayo2013, Kuzmin2019}. In \cite{Ashkenazy1982}, a microstrip antenna array was designed via a modular approach, where multiple antenna modules make up a large array to achieve higher gain. In \cite{Mayo2013}, a modular design of phased array antenna
was exploited to reduce the array size and achieve conformal capability. In \cite{Kuzmin2019}, a ring concentric antenna array that is composed of circular rotatable sub-arrays was designed in a modular manner, which is practically economic. However, all the aforementioned works mainly focused on the antenna designs instead of their communication channel modelling and performance. To fill this gap, modular massive MIMO (mmMIMO) system was studied in \cite{TWirth2020}, where modules each with a few antennas were utilized to construct a large full-dimension MIMO (FD-MIMO) system. In \cite{Bertilsson2018}, the authors presented a flexible modular architecture that accommodates
multiple antennas connected by a node, and further analysed its distributed computation complexity. By considering mmMIMO architecture with distributed antenna modules, the authors in \cite{Jeon2021} showed that the average throughput for multi-user systems could be improved than the conventional FD-MIMO. In our previous work \cite{Li2022}, we investigated wireless communications with modular XL-array, for which near-field NUSW channel model was developed, and the closed-form expression for the maximal achievable SNR was derived. However, this work only focused on one-dimensional (1D) modular XL-ULA architecture.\par
In this work, we extend our previous study on 1D modular XL-ULA \cite{Li2022} to the more general two-dimensional (2D) modular XL-array architecture. By accurately taking into account the variations of signal phase, amplitude and projected aperture across array elements, a near-field NUSW channel model for 2D modular XL-array is proposed, and the closed-form expression for its achievable SNR is then derived. We show that our developed near-field modelling and performance analysis include those for the existing collocated XL-array and far-field UPW model  as special cases. The main contributions of this paper are summarized as follows:\par
\begin{itemize}
\item First, we present the mathematical model of modular XL-array architecture, for which XL-array is arranged in a modular manner, and different modules are separated by distances  much larger than the signal wavelength to cater to the actual mounting structure. With the generic near-field NUSW model, a closed-form expression of the maximum achievable SNR with the optimal maximal-ratio combining (MRC) beamforming is derived. The result shows that the achievable SNR depends on the geometric parameters of modular XL-array, including the total planar array size, module separation distances along each dimension and the user's 3D location. Besides, for the special case with user located on the front of array center, i.e., the positive $x$-axis, the SNR can also be expressed in terms of the geometric angles formed by the user location and the array, which are termed \textit{horizontal angular span} and \textit{vertical angular span}, respectively \cite{Lu2021}.\par

\item Next, we show that our newly derived closed-form SNR expression includes the result for the conventional collocated XL-array as a special case. Furthermore, we study the asymptotic performance limit of modular XL-array. It is revealed that as the array size increases by adding more modules along each of the two dimensions, the corresponding SNRs approach to different upper bounds, which depend on the module spacing and effective aperture.  Besides, by considering the special case of the far-field UPW assumption, we show that the corresponding SNR depends on the projected aperture of the total array, which is usually ignored in the existing far-field model.\par
\item
    Furthermore, to obtain more insights, we consider the special case of 1D modular XL-ULA architecture. The result shows that the simplified SNR for the modular XL-ULA is governed by geometric angles, called \textit{angular span} and \textit{angular difference} \cite{Lu2021}.
    Finally, extensive numerical results are provided to demonstrate the importance of near-field NUSW modelling for modular XL-array communications.
\end{itemize}

The rest of this paper is organized as follows. Section
II introduces the mathematical modelling for modular XL-array. In Section III, the closed-form SNR
expression for modular XL-array is derived, and its performance scaling is studied. Next,
numerical results are presented in Section IV. Finally, we conclude this paper in Section V.

\emph{Notations}:
$||\cdot||$ represents the Euclidean norm and $|\cdot|$ denotes the absolute value of a complex number. The distribution of a circularly symmetric complex Gaussian vector with mean vector $\boldsymbol 0$ and covariance matrix ${\bf \Sigma} $ is denoted by ${\cal CN}({\boldsymbol 0},{\bf \Sigma})$. $\mathbb{C}^{M\times N}$ denotes the space of $M\times N$ complex-valued matrices.

\section{System Model} \label{model}\vspace{-1pt}

\begin{figure}[t]
\begin{centering}
\includegraphics[scale=0.65]{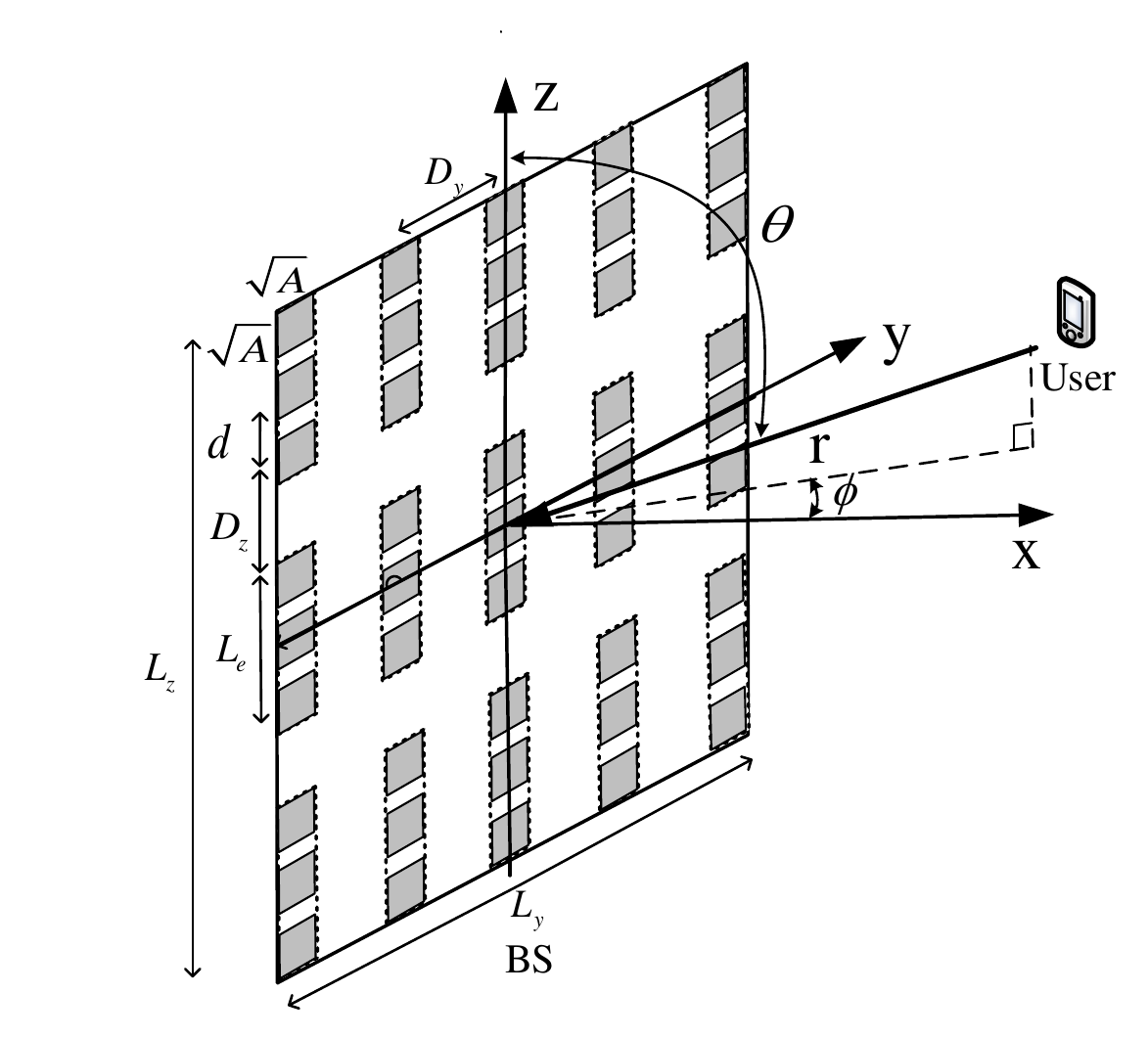}
\vspace{-0.3cm}
\caption{Wireless communication with a 2D modular XL-array.} \label{picture1}
\end{centering}
\vspace{-0.5cm}
\end{figure}

As shown in Fig. 2, we consider a wireless communication system,
where a base station (BS) equipped with a modular XL-array communicates with a single-antenna user. For ease of exposition, we assume that each module is of ULA architecture.
Without loss of generality, the modular XL-array is placed on the $y$-$z$ plane and centered at the origin.
The total number of modules is $N=N_y N_z$, where $N_y$ and $N_z$ are the number of modules along the $y$-axis and $z$-axis, respectively. Each ULA module consists of $M$ antenna elements with inter-element spacing denoted by $d$, which is on the wavelength scale, e.g., $d=\frac{\lambda}{2}$, with $\lambda$ denoting the signal wavelength. Furthermore, let $D_y=K_yd$ and $D_z=K_zd$ denote the module separations along the $y$-axis and $z$-axis, respectively, where $K_y\geq 1$ and $K_z\geq 1$ are integers.
For the special case when $K_y=1$ and $K_z=1$, the modular array architecture reduces to the conventional collocated UPA architecture \cite{Zeng2021}.
With the above notations, the physical length of each ULA module is thus $L_e=(M-1)d$, and
the physical dimensions of the modular XL-array along $y-$axis and $z-$axis are $L_y= K_y(N_y-1)d$ and $L_z=[K(N_z-1)+(M-1)]d$, respectively, with $K=M+K_z-1$.\par
Different from the conventional antenna array modelling in which each
antenna element is viewed as a sizeless point, we explicitly take the element
size into account, which is denoted as $\sqrt{A}\times \sqrt{A}$, where $\sqrt A \leq d$ \cite{Zeng2021,Bj2020}.
The effective aperture of each antenna element is $A_e=eA$,
where $0<e\le 1$ denotes the \textit{aperture efficiency}. Note that
for the hypothetical isotropic antenna element, we have $A_e=eA=\frac{\lambda^2}{4\pi}$.
Denote by $\xi \triangleq \frac{A}{d^2}\le 1$ the \textit{array occupation ratio},
which signifies the fraction of the total area of each module that is occupied by the antenna elements \cite{Zeng2021}.\par
For notational convenience, we assume that $N_y$, $N_z$ and $M$ are odd numbers.
Therefore, the central position of the $m$th antenna element of module $(n_y, n_z)$,
where $n_y= 0,\pm 1,...,\pm \frac{N_y-1}{2}$, $n_z= 0,\pm 1,...,\pm \frac{N_z-1}{2}$, and $m= 0,\pm 1,...,\pm \frac{M-1}{2}$, can be written as ${\bf w}_{n_y,n_z, m}=[0,y_{n_y},z_{n_z,m}]^T$,
where $y_{n_y}=n_yK_yd$ and $z_{n_z, m}=n_z[D_z+(M-1)d]+md=(Kn_z+m)d$.\par
Let $r$ denote the distance of the user from the center of the modular XL-array, and
$\theta \in [0,\pi]$ and $\phi \in [-\frac{\pi}{2},\frac{\pi}{2}]$ denote the zenith and azimuth angles, respectively. Therefore, the user's location can be written as ${\bf q}=[r\Psi,r\Phi,r\Omega]^T$,
with $\Psi \triangleq \sin\theta\cos\phi$, $\Phi \triangleq \sin\theta\sin\phi$
and $\Omega \triangleq \cos\theta$. Hence,
the distance between the user and the $m$th antenna element of module $(n_y,n_z)$ is expressed as
\begin{equation}\label{EQU-1} \vspace{-3pt}
\begin{split}
r_{n_y,n_z,m}&=||{\bf q}-{\bf w}_{n_y,n_z,m}||\\
&=r\sqrt{1-2m\epsilon\Omega-2Kn_z\epsilon\Omega-2K_yn_y\epsilon\Phi
+(m^2+K^2n_z^2+2Kmn_z+K_y^2n_y^2)\epsilon^2},
\end{split}
\end{equation}
where we have defined $\epsilon \triangleq \frac{d}{r}$.
Note that since antenna separation $d$ is on the wavelength scale,
we have $\epsilon \ll 1$.\par
Let $S_{n_{y}, n_{z}, m}=\left[n_yK_yd-\frac{\sqrt{A}}{2}, n_yK_yd+\frac{\sqrt{A}}{2}\right]
\times \left[(Kn_z+m)d-\frac{\sqrt{A}}{2}, (Kn_z+m)d+\frac{\sqrt{A}}{2}\right]$
denote the surface region of the $m$th antenna element of module $(n_y,n_z)$.
We focus on the basic free-space line-of-sight (LoS) propagation, for which
the channel power gain between the user and the $m$th antenna element of module $(n_y,n_z)$ can be written as \cite{Zeng2021, Bj2020}
\begin{equation}\label{EQU-2} \vspace{-3pt}
\bar{g}_{n_y, n_z, m}(r, \theta, \phi)=
\int_{S_{n_y, n_z, m}}{ \underbrace{\frac{1}{4 \pi\|\mathbf{q}-\mathbf{s}\|^{2}}}_{\text {Free-space path loss }} \underbrace{\frac{(\mathbf{q}-\mathbf{s})^{T} \hat{\mathbf{u}}_{x}}{\|\mathbf{q}-\mathbf{s}\|}}_{\text {Projection to signal direction }} d \mathbf{s},}
\end{equation}
where $\hat{\mathbf{u}}_{x}$ denotes the unit vector along the positive $x$-axis direction that is the normal direction of each array element. In practice, the size of each array element $A$ is on the wavelength scale, and hence
the variations of the wave propagation distance $||\bf{q}-\bf{s}||$ and signal direction
$\frac{\bf{s}-\bf{q}}{||\bf{q}-\bf{s}||}$ within each element are negligible. Then, we have
\begin{equation}\label{EQU-3} \vspace{-3pt}
\begin{split}
&\frac{1}{4 \pi\|\mathbf{q}-\mathbf{s}\|^{2}} \approx \frac{1}{4 \pi\left\|\mathbf{q}-\mathbf{w}_{n_y, n_z, m }\right\|^{2}}, \\
&\frac{(\mathbf{q}-\mathbf{s})^{T} \hat{\mathbf{u}}_{x}}{\|\mathbf{q}-\mathbf{s}\|} \approx \frac{\left(\mathbf{q}-\mathbf{w}_{n_y, n_z, m}\right)^{T} \hat{\mathbf{u}}_{x}}{\| \mathbf{q}-\mathbf{w}_{n_y, n_z, m} \|}, \forall \mathbf{s} \in S_{n_y, n_z, m}.
\end{split}
\end{equation}\par
As a result, the channel power gain can be expressed as
\begin{equation}\label{EQU-4} \vspace{-3pt}
\begin{split}
g_{n_y, n_z, m}(r, \theta, \phi)
&\approx \frac{1}{4 \pi\left\|\mathbf{q}-\mathbf{w}_{n_y, n_z, m}\right\|^{2}} \underbrace{eA \frac{\left(\mathbf{q}-\mathbf{w}_{n_y, n_z, m}\right)^{T} \hat{\mathbf{u}}_{x}}{\left\|\mathbf{q}-\mathbf{w}_{ n_y, n_z, m}\right\|}}_{\text {Element projected aperture }}
=\frac{eA r \sin \theta \cos \phi}{4 \pi\left\|\mathbf{q}-\mathbf{w}_{n_y, n_z, m}\right\|^{3}} \\
&=\frac{e\xi \epsilon^{2} \Psi}{4 \pi\left[1-2m\epsilon\Omega-2Kn_z\epsilon\Omega-2K_yn_y\epsilon\Phi
+(m^2+K^2n_z^2+2Kmn_z+K_y^2n_y^2)\epsilon^2\right]^{\frac{3}{2}}}.
\end{split}
\end{equation}\par
The array response vector for the user with the distance $r$ and direction $(\theta, \phi)$, denoted as ${\bf a}(r,\theta,\phi) \in \mathbb{C}^{(NM)\times 1}$, is expressed as
\begin{equation}\label{EQU-5} \vspace{-3pt}
\begin{split}
a_{n_y, n_z, m}(r, \theta, \phi)=\sqrt{g_{n_y, n_z, m}(r, \theta, \phi)} e^{-j \frac{2 \pi}{\lambda} r_{n_y, n_z, m}}.
\end{split}
\end{equation}\par
We focus on the uplink communication, while the results can be extended to the downlink scenario.
The received  signal at the BS after antenna beamforming is given by
\begin{equation}\label{EQU-6} \vspace{-3pt}
{y}={\bf v}^H{\bf a}(r,\theta, \phi)\sqrt{P}s+{\bf v}^H{\bf z},
\end{equation}
where ${\bf v}\in\mathbb{C}^{(N M)\times 1}$ is the receive beamforming vector with $||{\bf v}||=1$, $P$ is the transmit power,
$s$ is the transmitted signal of the user, and ${\bf z}$ denotes the additive white Gaussian noise (AWGN) that follows
the distribution of a circularly symmetric complex Gaussian vector with mean vector $\bf 0$ and covariance matrix ${\sigma}^2 {\bf I}_{N M}$, denoted as
${\bf z}\sim{\cal CN}({\bf 0},{\sigma}^2 {\bf I}_{N M})$. \par
Therefore, the resulting signal-to-noise ratio (SNR) can be written as
\begin{equation}\label{EQU-7} \vspace{-3pt}
\gamma_{\rm NUSW}=\bar P|{\bf v}^H {{\bf a}(r,\theta, \phi)}|^2,
\end{equation}
where $\bar P=\frac{P}{\sigma^2}$ is the transmit SNR.
With the optimal MRC receive beamforming ${\bf v}^*=\frac{{\bf a}(r,\theta, \phi)}{||{{\bf a}(r,\theta,\phi)}||}$,
the resulting maximum SNR is $\gamma_{\rm NUSW}=\bar P||{\bf a}(r,\theta, \phi)||^2$.
After substituting ${\bf a}(r,\theta,\phi)$ in (5), 
the maximum SNR can be written as
\begin{small}
\begin{equation}\label{EQU-8} \vspace{-3pt}
\begin{split}
&\gamma_{\rm NUSW}=\frac{\bar{P} eA \Psi}{4 \pi r^{2}}\times\\
&\sum_{n_{y}=-\frac{N_{y}-1}{2}}^{\frac{N_{y}-1}{2}}
\sum_{n_{z}=-\frac{N_{z}-1}{2}}^{\frac{N_{z}-1}{2}}
\sum_{m=-\frac{M-1}{2}}^{\frac{M-1}{2}}\frac{1}{\left[1-2m\epsilon\Omega-2Kn_z\epsilon\Omega-2K_yn_y\epsilon\Phi
+(m^2+K^2n_z^2+2Kmn_z+K_y^2n_y^2)\epsilon^2\right]^{\frac{3}{2}}}.
\end{split}
\end{equation}
\end{small}
\section{Closed-form SNR and performance analysis} \label{model}\vspace{-1pt}
In this section,
we first derive the closed-form expression for the maximum SNR in \eqref{EQU-8},
and then obtain the simplified SNR result when $(\theta,\phi)=(\frac{\pi}{2},0)$. We further study the special cases of the derived
expression, including the collocated XL-array architecture, asymptotic SNR limit, far-field assumption as well as 1D modular XL-ULA.
\subsection{Closed-form SNR} \label{model}\vspace{-1pt}
\begin{theo}\label{theo1}
For modular XL-array communications,
the maximum SNR in \eqref{EQU-8} can be obtained in closed-form as
\begin{equation}\label{EQU-9} \vspace{-3pt}
\begin{split}
\gamma_{\rm NUSW}&\approx\frac{e\xi\bar{P}d r\Psi }{4 \pi D_y[D_z+(M-1)d]}
\left[F\left(\frac{\tilde{L}_y}{2r} -\Phi, \frac{\tilde{L}_z}{2r}-\Omega\right)-\right.
F\left(\frac{\tilde{L}_y}{2r}-\Phi, \frac{\hat{L}_z}{2r}-\Omega\right)\\
&+F\left(\frac{\tilde{L}_y}{2r}-\Phi, \frac{\tilde{L}_z}{2r}+\Omega
\right)-
F\left(\frac{\tilde{L}_y}{2r}-\Phi,\frac{\hat{L}_z}{2r}+\Omega \right)+F\left(\frac{\tilde{L}_y}{2r}+\Phi, \frac{\tilde{L}_z}{2r}-\Omega
\right)\\
&-F\left(\frac{\tilde{L}_y}{2r}+\Phi,
\frac{\hat{L}_z}{2r}-\Omega \right)
+F\left(\frac{\tilde{L}_y}{2r}+\Phi, \frac{\tilde{L}_z}{2r}+\Omega
\right)-\left.F\left(\frac{\tilde{L}_y}{2r}+\Phi,
\frac{\hat{L}_z}{2r}+\Omega \right)
 \right],\\
\end{split}
\end{equation}
where $F(x,y)\triangleq \operatorname{arcsinh}\left(\frac{x}{\sqrt{\Psi^2+y^2}}\right)
+\frac{y}
{\Psi}
\arctan\left(\frac{xy}
{\Psi\sqrt{\Psi^2+x^2+y^2}} \right)$,
 $\tilde{L}_y=K_yN_yd=L_y+D_y$, $\tilde{L}_z=(KN_z+M)d=L_z+Md+D_z$, $\tilde{L}_e=Md=L_e+d$, $\hat{L}_z=(KN_z-M)d=\tilde{L}_z-2\tilde{L}_e$, and $\operatorname{arcsinh}(x)\triangleq\ln(x+\sqrt{1+x^2})$ is hyperbolic arcsine function.
Note that $\tilde{L}_y\approx L_y$ and $\tilde{L}_z\approx L_z$ hold when $N_y\gg1$ and $N_z\gg1$, $\tilde{L}_e\approx L_e$ is true when $M\gg 1$, and the term $\frac{e\xi\bar{P}d r\Psi }{4 \pi D_y[D_z+(M-1)d]}$ outside the bracket of \eqref{EQU-9} can be also rewritten as $\frac{e\xi\bar{P} r\Psi }{4 \pi D_yK}$ by letting $D_z=K_zd$.
\end{theo}
\begin{pproof}Please refer to Appendix A. $\hfill \blacksquare$\end{pproof}  \par
Theorem 1 shows that the maximum SNR for modular XL-array communications depends on
the geometric parameters, including the total planar array size along $y$-axis and $z$-axis, i.e., $\tilde{L}_y$ and $\tilde{L}_z$, the corresponding module separation distances along $y$-axis and $z$-axis, i.e., $D_y$ and $D_z$, as well as the user's location $\bf q$, while irrespective of the individual element parameters, including the element size $A$ and the inter-element spacing $d$. This is expected since the inter-element spacing $d$ is much smaller than the link distance $r$, so that the
sum form \eqref{EQU-8} can be well approximated by the integration result \eqref{EQU-9}. Besides, it is observed that the SNR in \eqref{EQU-9} is also determined by the array occupation ratio of each module $e\xi$ and the user's projected distance along $x$-axis $r\Psi$, which account for the impact of effective projected aperture. Furthermore, the terms inside the bracket of \eqref{EQU-9} indicate the variation of the total planar array size $\tilde{L}_y$ and $\tilde{L}_z$, and the user's projected distance to the array $r\Phi$ and $r\Omega$.
Moreover, the closed-form expression \eqref{EQU-9} is applicable for different practical mounting structures by flexibly adjusting the module separation distances $D_y$ and $D_z$, or the number of modules $N$ and antenna elements in each module $M$.\par

\begin{figure}[t]
\begin{centering}
\includegraphics[scale=0.58]{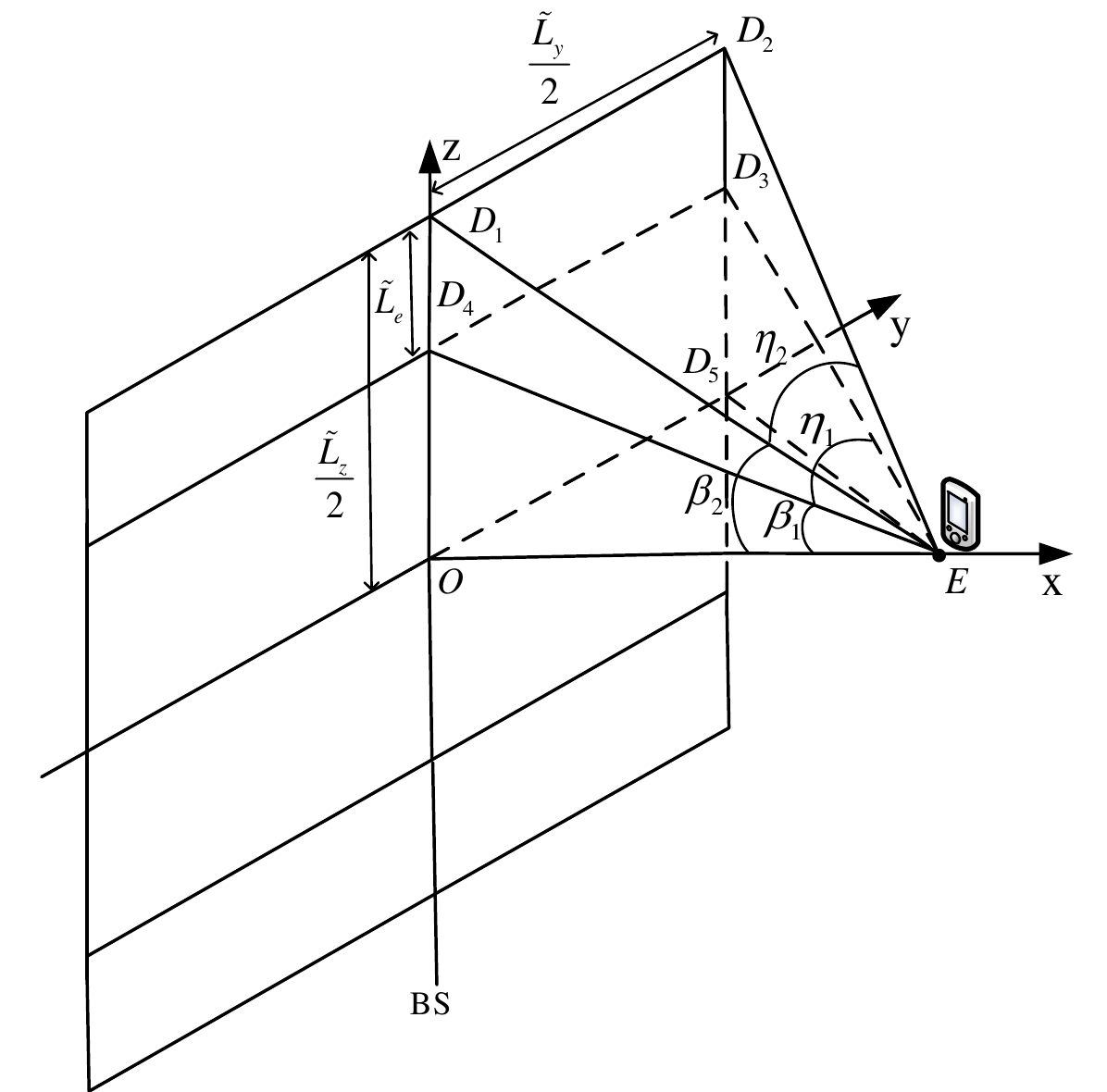}
\vspace{-0.3cm}
\caption{The geometrical relationships when the user is located on the $x$-axis.} \label{picture1}
\end{centering}
\vspace{-0.5cm}
\end{figure}

\subsection{Special Case with User Located on $x$-axis} \label{model}\vspace{-1pt}
To gain useful insights from the closed-form expression \eqref{EQU-9}, we consider the following special case.
\begin{lemma}\label{lemma2}
When the user is located on the $x$-axis, i.e., $\theta=\frac{\pi}{2}$ and $\phi=0$,
the SNR expression \eqref{EQU-9} reduces to
\begin{equation}\label{EQU-16} \vspace{-3pt}
\gamma_{\rm NUSW}\left(\theta=\frac{\pi}{2},\phi=0\right)=\frac{e\xi \bar{P}d r}{\pi D_y[D_z+(M-1)d]}
\left[F_1\left(\frac{\tilde{L}_y}{2r}, \frac{\tilde{L}_z}{2r}\right)-F_1\left(\frac{\tilde{L}_y}{2r},
\frac{\hat{L}_z}{2r} \right)
 \right],
\end{equation}
where $F_1(x,y)\triangleq \operatorname{arcsinh}\left(\frac{x}{\sqrt{1+y^2}}\right)
+y\arctan\left(\frac{xy}{\sqrt{1+x^2+y^2}} \right)$.
\end{lemma}\par
As illustrated in Fig. 3, denote by $D_1\sim D_5$ the vertices of one quarter of modular XL-array. We have $D_1D_2=D_3D_4=OD_5=\frac{\tilde{L}_y}{2}$, $OD_1=D_2D_5=\frac{\tilde{L}_z}{2}$ and
$D_1D_4=D_2D_3=\tilde{L}_e$.
An alternative expression of \eqref{EQU-16} can be equivalently written as
\begin{equation}\label{EQU-17} \vspace{-3pt}
\begin{split}
&\gamma_{\rm NUSW}\left(\theta=\frac{\pi}{2},\phi=0\right)=\frac{e\xi \bar{P} dr}{\pi D_y[D_z+(M-1)d]}
\left[\operatorname{arcsinh}\left(\tan\eta_2\right)\right.\\
&\left.+\tan\beta_2
\arctan\left(\tan\beta_2\sin\eta_2 \right)-\operatorname{arcsinh}\left(\tan\eta_1\right)-\tan\beta_1
\arctan\left(\tan\beta_1\sin\eta_1 \right)
 \right],\\
\end{split}
\end{equation}
where $\eta_1=\arctan\frac{\frac{\tilde{L}_y}{2}}{\sqrt{r^2+\left(\frac{\hat{L}_z}{2}\right)^2}}$,
$\eta_2=\arctan\frac{\frac{\tilde{L}_y}{2}}{\sqrt{r^2+\left(\frac{\tilde{L}_z}{2}\right)^2}}$, $\beta_1=\arctan\frac{\hat{L}_z}{2r}$, and $\beta_2=\arctan\frac{\tilde{L}_z}{2r}$.
The expression \eqref{EQU-17} is used to interpret the obtained SNR result in \eqref{EQU-16} in terms of geometrical relationships. Specifically, the obtained SNR
depends on \textit{horizontal angular spans}, i.e., $\eta_1$, $\eta_2$, and \textit{vertical angular spans}, i.e., $\beta_1$ and $\beta_2$. As horizontal angular span $\eta_2$ or vertical angular span
$\beta_2$ increases, the term $\operatorname{arcsinh}\left(\tan\eta_2\right)+\tan\beta_2
\arctan\left(\tan\beta_2\sin\eta_2 \right)$ will become large, leading to the SNR increasing to a higher value.
Similarly, the increase of horizontal angular span $\eta_1$ or vertical angular span $\beta_1$ will lead to smaller SNR.\par

\subsection{Degeneration to Collocated XL-array} \label{model}\vspace{-1pt}
For the special case when the module separations $D_y$ and $D_z$ are equal to inter-antenna spacing $d$, the SNR of modular XL-array given in \eqref{EQU-9} degenerates to that of conventional collocated XL-array \cite{Zeng2021}, as shown in Corollary 2.
\begin{lemma}\label{theo1}
When $D_y=D_z=d$,
the SNR in (9) for modular XL-array communication reduces to
\begin{equation}\label{EQU-18} \vspace{-3pt}
\begin{split}
\gamma_{\rm NUSW}&=\frac{e\xi \bar{P}}{4 \pi }\left[G\left(\frac{N_yd}{2 r}-\Phi, \frac{N_zMd}{2 r}-\Omega\right)+G\left(\frac{N_yd}{2 r}-\Phi, \frac{N_zMd}{2 r}+\Omega\right)\right.\\
&+G\left(\frac{N_yd}{2 r}+\Phi, \frac{N_zMd}{2 r}-\Omega\right)\left.+G\left(\frac{N_yd}{2 r}+\Phi, \frac{N_zMd}{2 r}+\Omega\right)\right],
\end{split}
\end{equation}
where $G(x, y) \triangleq \arctan \left(\frac{x y}{\Psi \sqrt{\Psi^{2}+x^{2}+y^{2}}}\right)$.
\end{lemma}\par
\begin{pproof}Please refer to Appendix B. $\hfill \blacksquare$\end{pproof}  \par
An useful observation of \eqref{EQU-18} is that the terms inside the
bracket account for the impact of the variation of the total planar array size $N_yd$ and $N_zMd$, and the user's projected distance to the array $r\Phi$ and $r\Omega$. It is also observed from Corollary 2 that compared with the obtained SNR for collocated XL-array that
only depends on the planar array size $N_yd$ and $N_zMd$ \cite{Zeng2021},  that for modular XL-array is affected by the geometric
parameters, including the planar physical size $L_y$ and $L_z$, as well as module separations along different dimensions $D_y$ and $D_z$. As a result, the conventional collocated XL-array which needs extremely large continuous surface is typically limited by the actual mounting structure, while module separations along different dimensions $D_y$ and $D_z$ render modular XL-array more flexible for deployment.\par
\subsection{Near-field Asymptotic SNR Scaling} \label{model}\vspace{-1pt}
As the array size of modular XL-array grows, the asymptotic limit of \eqref{EQU-9} is revealed in the following corollary, by considering three different cases.\par
\begin{lemma}\label{lemma2}
Case 1: As the number of modules along $z$-axis $N_z$ increases indefinitely while fixing $N_y$, we have
\begin{equation}\label{EQU-19} \vspace{-3pt}
\begin{split}
&\lim_{N_z \to \infty} \gamma_{\rm NUSW}=\frac{\overline{P}M eA}{2\pi D_y[(M-1)d+D_z]}
\left[\arctan\left(\frac{\tilde{L}_y-2r\Phi}{2r\Psi}\right)
+\arctan\left(\frac{\tilde{L}_y+2r\Phi}{2r\Psi}\right)\right];\\
\end{split}
\end{equation}\par
Case 2: As $N_y$ increases indefinitely while fixing $N_z$, we have\par
\begin{equation}\label{EQU-1911} \vspace{-3pt}
\begin{split}
&\lim_{N_y \to \infty} \gamma_{\rm NUSW}=
\frac{\overline{P}M eA}{2\pi D_y[(M-1)d+D_z]}
\left[\arctan\left(\frac{\tilde{L}_z-2r\Omega}{2r\Psi}\right)\right.
\left.+\arctan\left(\frac{\tilde{L}_z+2r\Omega}{2r\Psi}\right)\right]\\
&-\frac{\overline{P}M eAr\Psi}{\pi D_y[(M-1)d+D_z]}\left[\frac{\tilde{L}_e}{(\tilde{L}_z-\tilde{L}_e-2r\Omega)^2+(2r\Psi)^2}\right.\left.+\frac{\tilde{L}_e}{(\tilde{L}_z-\tilde{L}_e+2r\Omega)^2+(2r\Psi)^2}\right];\\
\end{split}
\end{equation}\par
Case 3: As both $N_y$ and $N_z$ increase indefinitely, we have
\begin{equation}\label{EQU-192} \vspace{-3pt}
\lim_{N_y, N_z \to \infty} \gamma_{\rm NUSW}=\frac{\overline{P}}{2D_y[(M-1)d+D_z]}\underbrace{M eA}_{\text {Module effective aperture}}.
\end{equation}

\end{lemma}
\begin{pproof} Please refer to Appendix C.$\hfill \blacksquare$ \end{pproof}\par

Corollary 3 shows that as the size of modular XL-array increases, the resulting SNR would approach to constant values, rather than increasing unboundedly. Such results are expected and comply with the law of energy conservation. Specifically, it is observed that the asymptotic SNR limits in (13) and (14) are
not only dependent on the module separation  distances  $D_y$ and $D_z$, and the effective aperture of each module, i.e., $MeA$,
 but also determined by the user's location $\bf q$ and the array size $\tilde{L}_y$ or $\tilde{L}_z$. Instead,
the asymptotic SNR limit in (15) only decreases with the module separation distances $D_y$ and $D_z$, while
increasing with the effective aperture of each module, i.e., $MeA$. Besides, for the special
case when $D_y=D_z=d$, \eqref{EQU-192} degenerates to the asymptotic SNR for collocated XL-array, i.e., $\gamma_{\rm collocated}=\frac{\overline{P}eA}{2d^2}$ \cite{Zeng2021}.
Furthermore, to clearly show the relationship between the SNR asymptotic limits of collocated and modular XL-array architectures, we define the following SNR ratio as
\begin{equation}\label{EQU-191} \vspace{-3pt}
\Gamma=\frac{\gamma_{\rm collocated}}{\lim_{N_y, N_z \to \infty} \gamma_{\rm NUSW}}=\frac{K_y(K_z+M-1)}{M},
\end{equation}
which is a constant value dependent on the module separations with $D_y=K_yd$ and $D_z=K_zd$ along each direction, as well as the number of elements $M$ in each module.\par
For the hypothetical isotropic array elements separated by half-wavelength, i.e.,
$A_e=eA=\frac{\lambda^2}{4\pi}$ and $d=\frac{\lambda}{2}$, \eqref{EQU-192} can be simplified as
\begin{equation}\label{EQU-20} \vspace{-3pt}
\lim_{N_y, N_z \to \infty} \gamma_{\rm NUSW}=\frac{\overline{P}M}{2\pi K_y(K_z+M-1)}.
\end{equation}

\subsection{Degeneration to Far-Field UPW Model} \label{model}\vspace{-1pt}
In the following corollary, we further study the far-field behaviour of the
generic SNR expression in \eqref{EQU-9}.\par
\begin{lemma}\label{lemma3}
When $r\Psi \gg \tilde{L}_y$, $r\Psi \gg \tilde{L}_z$ and $\frac{\frac{\Phi}{\Psi} \frac{\Omega}{\Psi}}{\sqrt{1+\left(\frac{\tilde{L}_y}{2 r \Psi} \pm \frac{\Phi}{\Psi}\right)^{2}+\left(\frac{\tilde{L}_z-\tilde{L}_e}{2 r \Psi} \pm \frac{\Omega}{\Psi}\right)^{2}}} \ll 1$, the SNR expression in (9) reduces to
\begin{equation}\label{EQU-21} \vspace{-3pt}
\gamma_{\rm NUSW}\approx \gamma_{\rm UPW}= \frac{\overline{P}}{4\pi r^2}\underbrace{N_yN_zMeA\Psi}_{\text {Total projected aperture}}.
\end{equation}
\end{lemma}
\begin{pproof} Please refer to Appendix D.$\hfill \blacksquare$ \end{pproof}\par
Corollary 4 shows that with the conventional far-field UPW assumption, the beamforming gain increases linearly and unboundedly with the total number of array elements, i.e., $N_yN_zM$, which is consistent with the SNR result under the UPW assumption in \cite{Zeng2021, Bj2020}. Corollary 4 also indicates that our derived SNR expression in \eqref{EQU-9} is general, since it is applicable to both
near-field and far-field scenarios. In addition,
it is observed from \eqref{EQU-21} that under the far-field UPW assumption, the resulting SNR only
depends on the projected aperture of the total antenna array, i.e., $N_yN_zMeA\Psi$, while being independent of the module separation distances $D_y$ and $D_z$. The reason behind the phenomenon is that the user's projected distance along $x$-axis $r\Psi$ is much longer than the array size $\tilde{L}_y$ and $\tilde{L}_z$, so that the effect of the module separation distances $D_y$ and $D_z$ can be ignored.\par
As a comparison, the SNR under the conventional far-field UPW assumption without considering the total projected aperture is
\begin{equation}\label{EQU-22} \vspace{-3pt}
\gamma_{\rm UPW}= \frac{\overline{P}N_yN_zM\beta_0}{r^2},
\end{equation}
where $\beta_0$ is the channel gain at the reference distance of $r_0=1$ $m$. For the isotropic array element, i.e., $A_e=eA=\frac{\lambda^2}{4\pi}$ and $\beta_0=(\frac{\lambda}{4\pi})^2$, \eqref{EQU-21} differs from \eqref{EQU-22} by $\Psi$.
This indicates
that the conventional far-field UPW model over-estimates the SNR value in \eqref{EQU-21} as the latter takes into account the
projected aperture since $\Psi \le 1$.\par

\begin{figure}[t]
\begin{centering}
\includegraphics[scale=0.7]{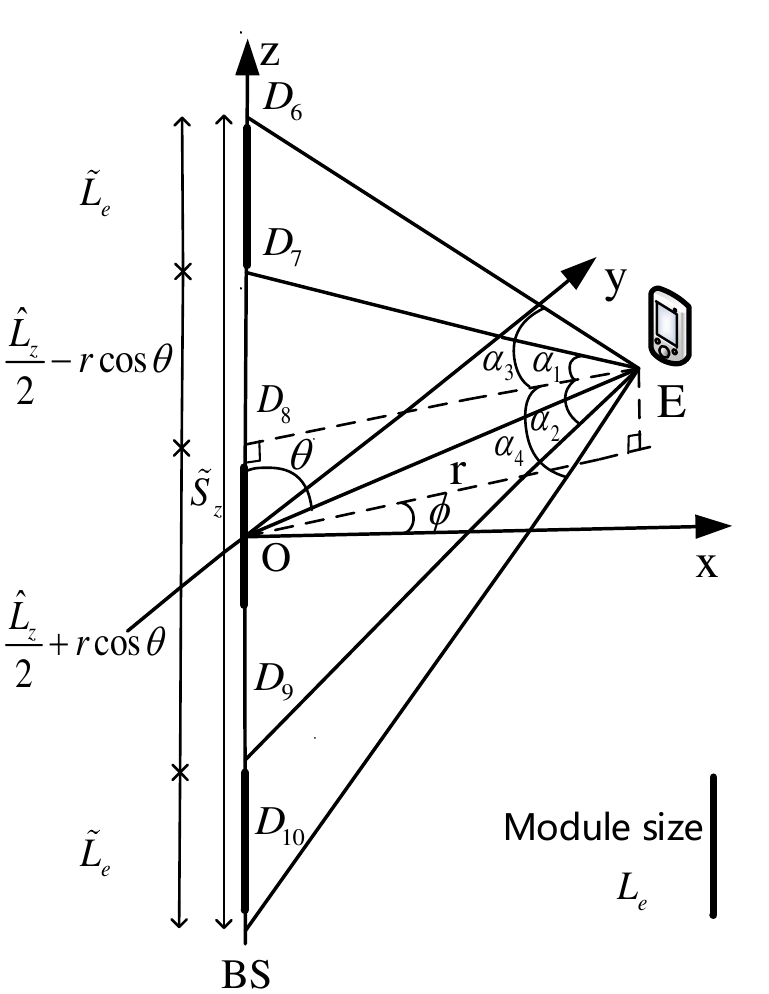}
\vspace{-0.3cm}
\caption{The geometrical relationships for 1D modular XL-ULA.} \label{picture1}
\end{centering}
\vspace{-0.5cm}
\end{figure}

\subsection{Special Case of 1D Modular XL-ULA} \label{model}\vspace{-1pt}

To obtain more insights of the closed-form expression in \eqref{EQU-9}, we consider the special case when $N_y=1$ and $K_y=1$, for which \eqref{EQU-9} reduces to the result of 1D modular XL-ULA.
\begin{lemma}\label{theo1}
For the special case of modular XL-ULA, i.e., $N_y=1$ and $K_y=1$,
the SNR expression in \eqref{EQU-9} degenerates to
\begin{equation}\label{EQU-23} \vspace{-3pt}
\begin{split}
\gamma_{\rm NUSW,1D}&=\frac{e\xi \bar{P}d\cos\phi}{4 \pi [D_z+(M-1)d]\sin\theta}
\left[H\left(\frac{\tilde{L}_z}{2r}-\Omega\right)-H\left(\frac{\hat{L}_z}{2r}-\Omega\right)\right.\\
&+H\left(\frac{\tilde{L}_z}{2r}+\Omega\right)-\left.H\left(\frac{\hat{L}_z}{2r}+\Omega\right)
\right],\\
\end{split}
\end{equation}
where $H(x) \triangleq \sqrt{\sin^2\theta+x^2}$.
\end{lemma}
\begin{pproof}Please refer to Appendix E. $\hfill \blacksquare$\end{pproof}  \par
Corollary 5 shows that the obtained SNR depends on the geometries of modular XL-ULA, such as the physical length $\tilde{L}_z$, the module separation $D_z$ and the user's location $\bf q$.
This result is also applicable to different practical discrete deployment surfaces via adjusting the module separation $D_z$ accordingly.
Furthermore, the resulting SNR can be expressed in an alternative form in terms of four geometric angles shown in Fig. 4, which can be expressed as


\begin{equation}\label{EQU-230} \vspace{-3pt}
\begin{split}
\gamma_{\rm NUSW,1D}&=\frac{e\xi \bar{P}d\cos\phi}{4 \pi [D_z+(M-1)d]}\left[\frac{\cos\alpha_3+\cos\alpha_4}{\cos\alpha_3\cos\alpha_4}-\frac{\cos\alpha_1+\cos\alpha_2}{\cos\alpha_1\cos\alpha_2}\right]\\
&=\frac{e\xi \bar{P}d\cos\phi}{\pi [D_z+(M-1)d]}\left[\frac{\cos(\frac{\Delta_{s,2}}{2})\cos(\frac{\Delta_{d,2}}{2})}{\cos(\Delta_{s,2})+\cos(\Delta_{d,2})}-\frac{\cos(\frac{\Delta_{s,1}}{2})\cos(\frac{\Delta_{d,1}}{2})}{\cos(\Delta_{s,1})+\cos(\Delta_{d,1})}\right],\\
\end{split}
\end{equation}
where $\Delta_{s,1}=\alpha_1+\alpha_2$, $\Delta_{d,1}=\alpha_2-\alpha_1$,  $\Delta_{s,2}=\alpha_3+\alpha_4$, $\Delta_{d,2}=\alpha_4-\alpha_3$, $\alpha_1=\arctan\left(\frac{\frac{\hat{L}_z}{2}-r\cos\theta}{r\sin\theta}\right)$,
$\alpha_2=\arctan\left(\frac{\frac{\hat{L}_z}{2}+r\cos\theta}{r\sin\theta}\right)$,
$\alpha_3=\arctan\left(\frac{\frac{\tilde{L}_z}{2}-r\cos\theta}{r\sin\theta}\right)$,
and $\alpha_4=\arctan\left(\frac{\frac{\tilde{L}_z}{2}+r\cos\theta}{r\sin\theta}\right)$.
Note that \eqref{EQU-230} can be easily shown by using the trigonometric functions.
This result indicates that \eqref{EQU-230} depends on the angular spans $\Delta_{s,1}$ and $\Delta_{s,2}$, as well as the angular differences $\Delta_{d,1}$ and $\Delta_{d,2}$.\par


\begin{case}\label{theo1}
The asymptotical SNR of modular XL-ULA as $N_z$ goes infinitely is given by
\begin{equation}\label{EQU-24} \vspace{-3pt}
\begin{split}
&\lim_{N_z \to \infty} \gamma_{\rm NUSW,1D}=\frac{\overline{P}\cos\phi}{2\pi [D_z+(M-1)d]r\sin\theta}\underbrace{MeA}_{\text {Module effective aperture}}.
\end{split}
\end{equation}
\end{case}
\begin{pproof}
Similar to the proof of Corollary 3, the asymptotical limit of \eqref{EQU-23} is obtained via the first-order Taylor series expansion for relatively small $\frac{Md}{2r}$.
$\hfill \blacksquare$\end{pproof}  \par

\begin{case}\label{theo1}
When $r\Psi \gg \tilde{L}_z$, the SNR of modular XL-ULA is in accordance with the SNR result based on the far-field UPW assumption, i.e.,
\begin{equation}\label{EQU-26} \vspace{-3pt}
\gamma_{\rm NUSW,1D}\approx \gamma_{\rm UPW,1D}= \frac{\overline{P}}{4\pi r^2}\underbrace{N_zMeA\Psi}_{\text {Total projected aperture}}.
\end{equation}
\end{case}
\begin{pproof}The proof of this lemma is similar to that of Corollary 4, which is omitted for brevity. $\hfill \blacksquare$\end{pproof}  \par

Lemma 2 shows that for modular XL-ULA, the SNR based on the NUSW model is applicable to both near-field and far-field scenarios. Besides,
compared to the conventional far-field UPW model that ignores the impact of projected
aperture, i.e., $\gamma_{\rm UPW,1D}= \frac{\overline{P}N_z M\beta_0}{r^2}$,
the asymptotical SNR in (23) also depends on the projected aperture, i.e., $N_zMeA\Psi$. However,
such two SNR results $\gamma_{\rm UPW,1D}$ are independent of the modular separation $D_z$, due to the UPW assumption that user's projected distance along $x$-axis is much longer than the array size, i.e., $r\Psi \gg \tilde{L}_z$.\par
\section{numerical results} \label{model}\vspace{-1pt}

In this section, numerical results are provided to
validate our obtained analytic results for modular XL-array communications. Unless otherwise specified,
the number of array elements in each module is $M=9$, the transmit SNR is $\overline P =90$ $\rm dB$, the aperture efficiency is $e=1$, the carrier frequency is $2.38$ GHz, and the inter-element spacing is $d=\frac{\lambda}{2}=0.0628$ $\rm m$.
Besides, the distance from the user to the center of modular XL-array is $r=25$ $\rm m$, and
the number of modules along the $y$-axis and $z$-axis are set as $N_y=64$ and $N_z=64$, respectively.
The module separations along the $y$-axis and $z$-axis are given by $D_y=D_z=10d=0.628$ $\rm m$. \par

\begin{figure}[t]
  \centering
  \subfigure[]{
    \label{1} 
    \includegraphics[scale=0.52]{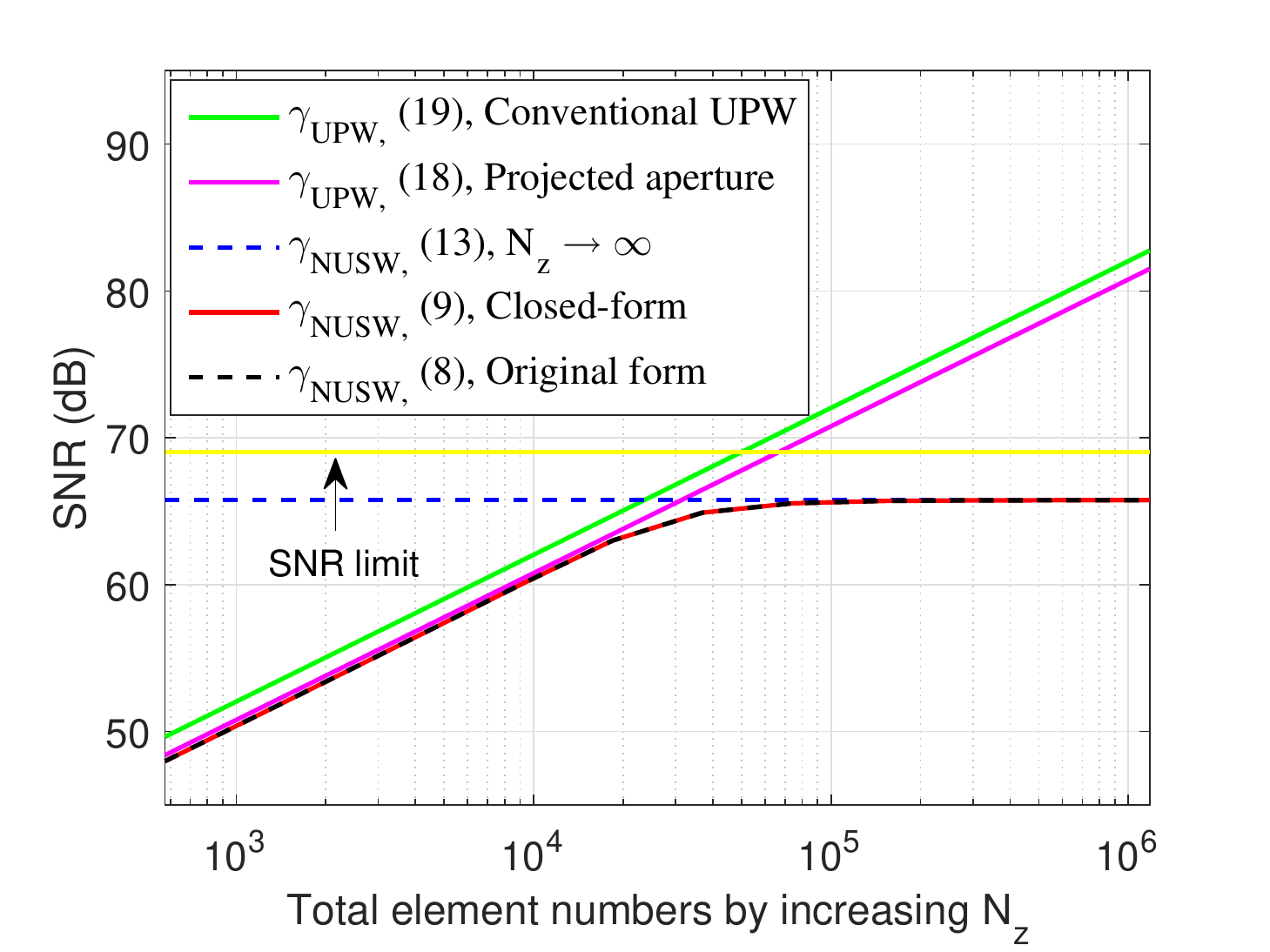}}
  \hspace{0in}
  \subfigure[]{
    \label{2}
    \includegraphics[scale=0.5]{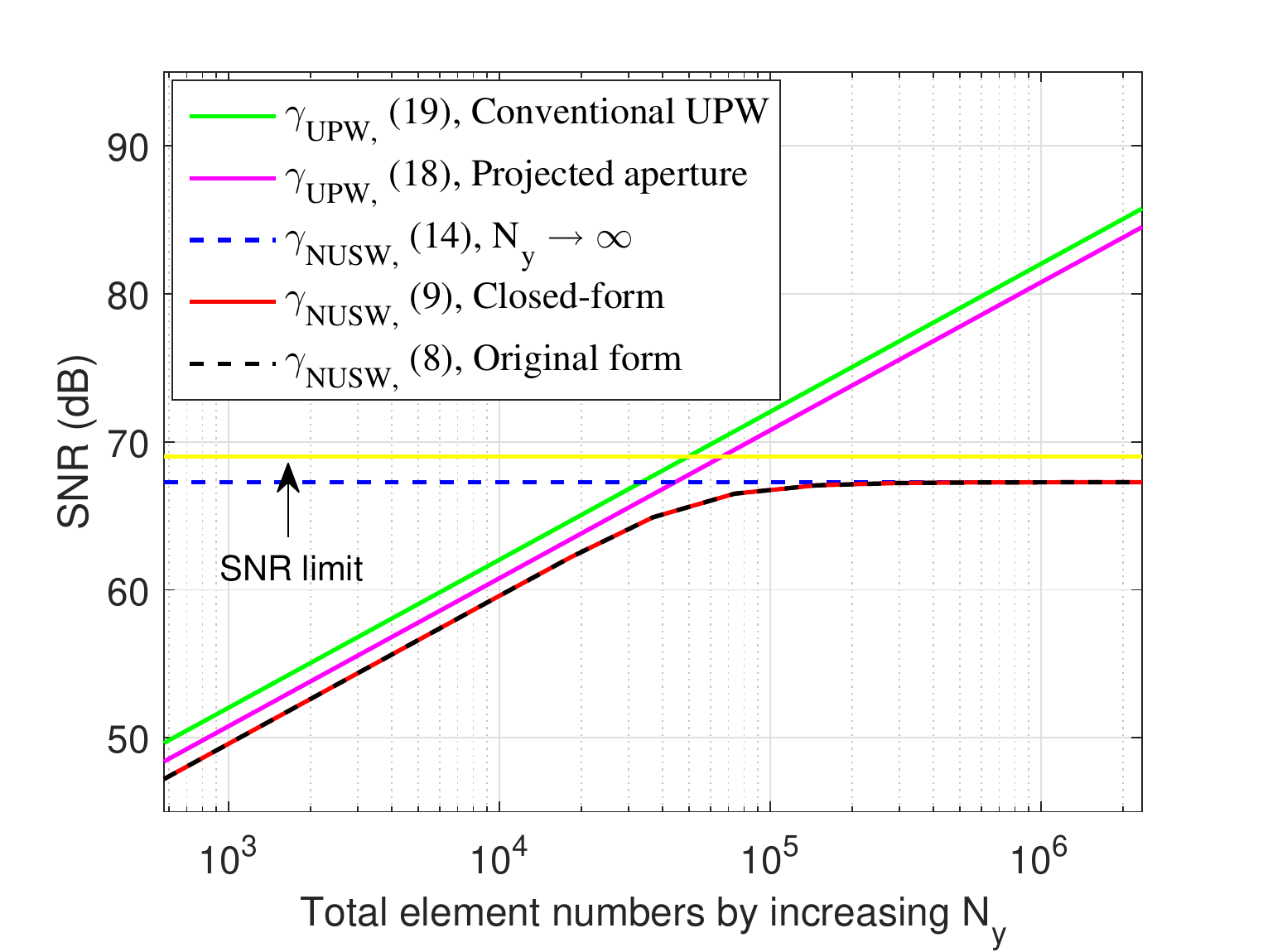}}
    \hspace{0in}
  \caption{SNRs for different models versus the total number of array elements, i.e., (a) Increasing $N_z$; (b) Increasing $N_y$.}\label{pic1}
  \label{12}
  \vspace{-0.5cm}
\end{figure}

\begin{figure}[t]
\begin{centering}
\includegraphics[scale=0.53]{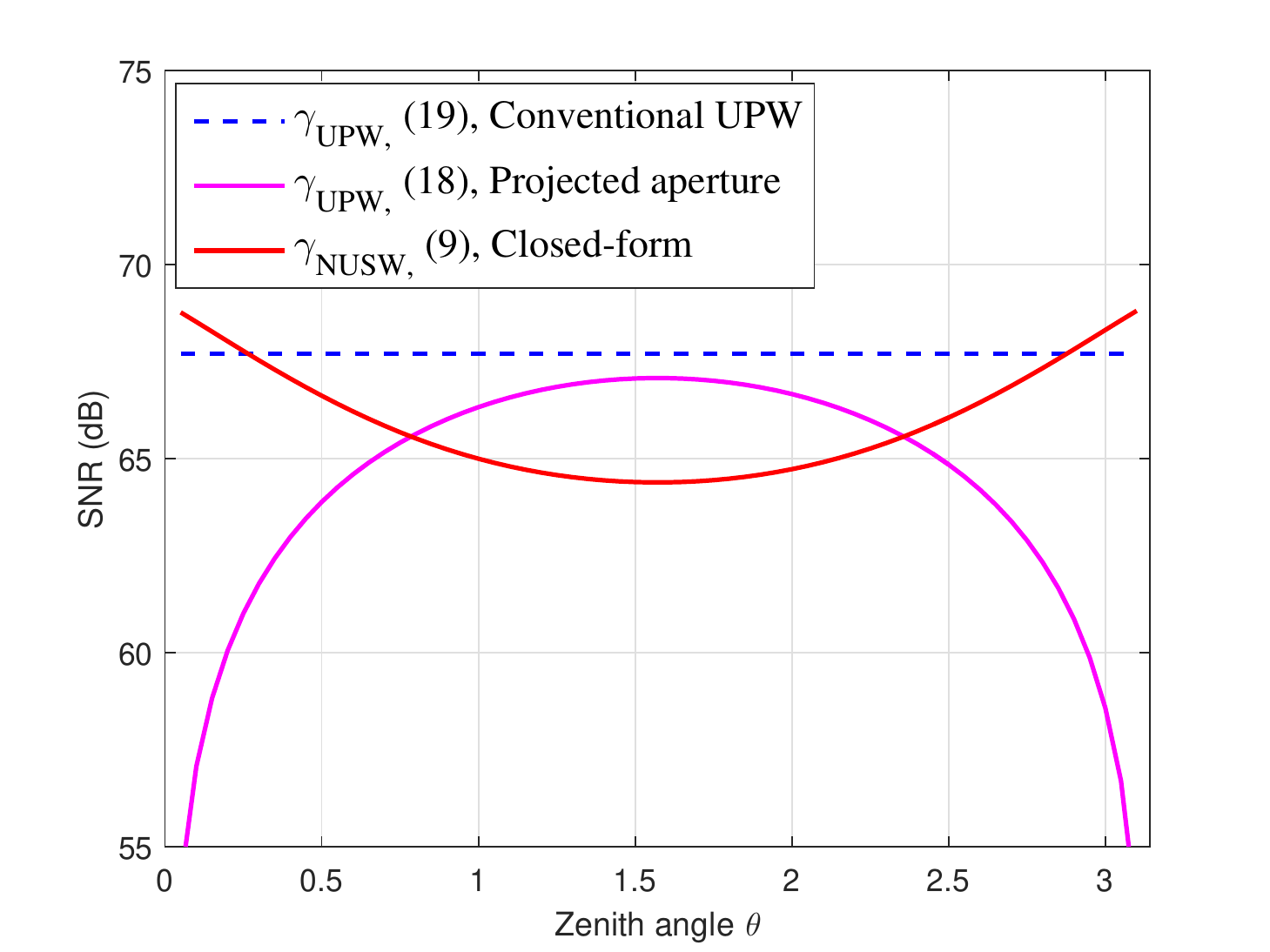}
\vspace{-0.3cm}
\caption{SNRs for different models versus zenith angle $\theta$.} \label{picture12}
\end{centering}
\vspace{-0.3cm}
\end{figure}
\begin{figure}[t]
\begin{centering}
\includegraphics[scale=0.53]{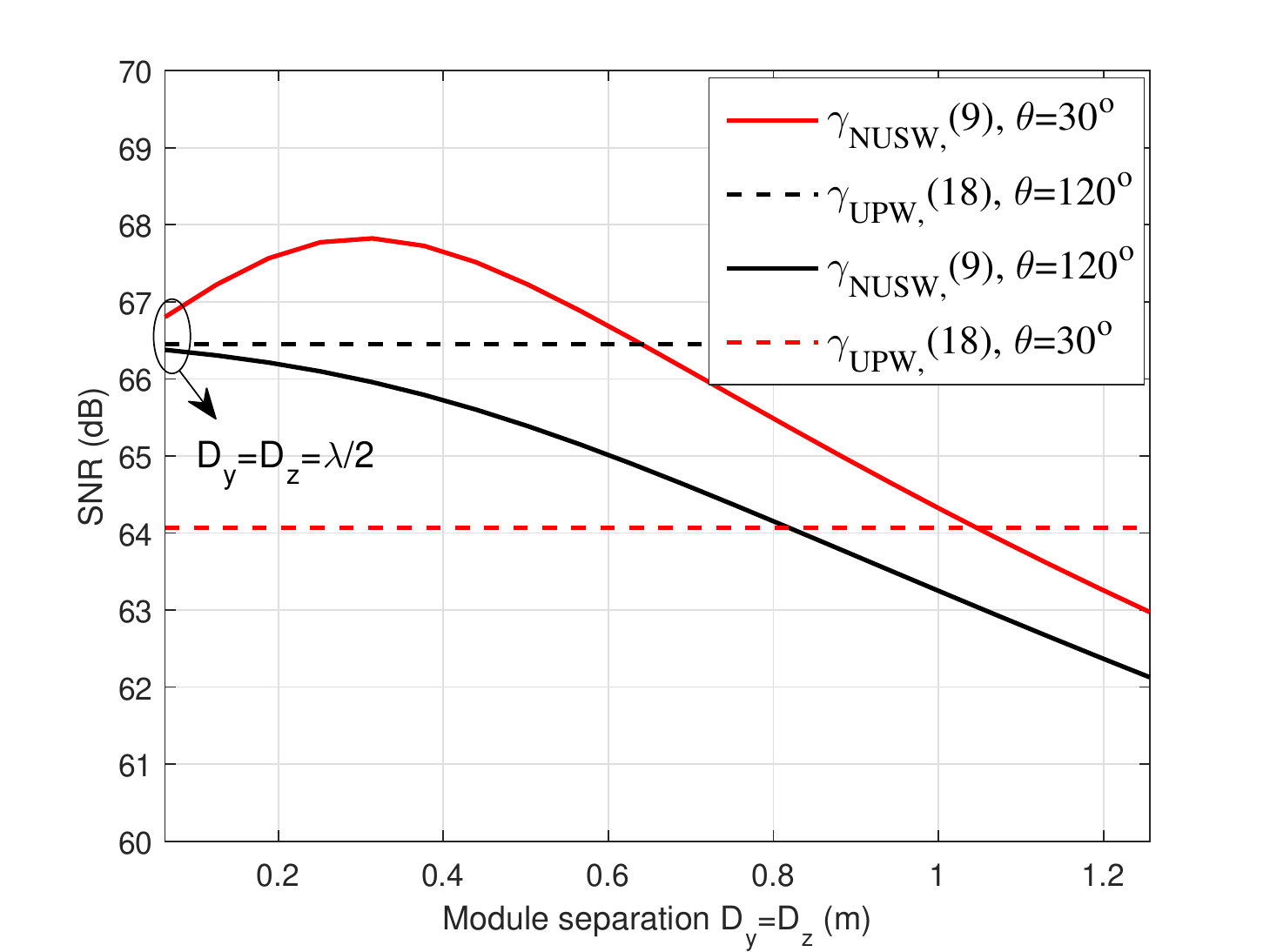}
\vspace{-0.3cm}
\caption{SNRs versus module separations $D_y=D_z$ with different zenith direction $\theta$.} \label{picture13}
\end{centering}
\vspace{-0.3cm}
\end{figure}

Fig.~\ref{pic1}(a) shows the SNRs $\gamma_{\rm NUSW}$ and $\gamma_{\rm UPW}$
versus the number of total array elements $N_yN_zM$, by increasing the number of modules $N_z$ along the $z$-axis. Note that the direction of the user is given as $(\theta,\phi)=(60^\circ,30^\circ)$ and the ``SNR limit" shown in the figure is obtained by letting $N_y, N_z\to\infty$, i.e., (15).
As can be seen from Fig.~\ref{pic1}(a),
the closed-form SNR expression in \eqref{EQU-9} perfectly matches with the original summation expression in \eqref{EQU-8},
which demonstrates the accuracy of Theorem 1. When the total number of array elements $N_yN_zM$ is relatively small,
both $\gamma_{\rm NUSW}$ and $\gamma_{\rm UPW}$ increase linearly with $N_yN_zM$, which is consistent with Corollary 4.
However, as $N_yN_zM$ becomes large by increasing $N_z$,
the SNR results based on NUSW and UPW models exhibit two different scaling laws, i.e.,
approaching to a constant value versus growing linearly unboundedly, which are in accordance with (13) and (18), respectively.
It is also observed that the SNR in (19) which ignores the variation of the projected aperture across array elements over-estimates the SNR obtained by (18). Similar to Fig.~\ref{pic1}(a), Fig.~\ref{pic1}(b) shows the SNRs $\gamma_{\rm NUSW}$ and $\gamma_{\rm UPW}$ versus total number of elements $N_yN_zM$ by increasing the module numbers $N_y$ along the $y$-axis. It is revealed that similar observations as Fig.~\ref{pic1}(a) can be obtained.
Moreover, it is also observed from Fig.~\ref{pic1} that the SNR limit of (15) with both $N_y$ and $N_z$ going to infinity is larger than its counterparts when only $N_y$ or $N_z$ grows, as expected.\par

Fig.~\ref{picture12} plots the SNRs $\gamma_{\rm NUSW}$ and $\gamma_{\rm UPW}$ versus zenith angle $\theta$ by fixing the azimuth angle $\phi=30^\circ$.
It is observed that for relatively small values of $\sin\theta$, e.g., $\theta=10^\circ$ or $\theta=170^\circ$, the SNR under UPW modelling in (18) under-estimates the result in (9), while the reverse is true as $\sin\theta$ is large, e.g., $\theta=90^\circ$.
It is worth noting that the UPW-based result in (19) without considering
the effect of projected aperture is independent of the user direction $\theta$, since isotropic array elements are assumed.\par

\begin{figure}[t]
\begin{centering}
\includegraphics[scale=0.46]{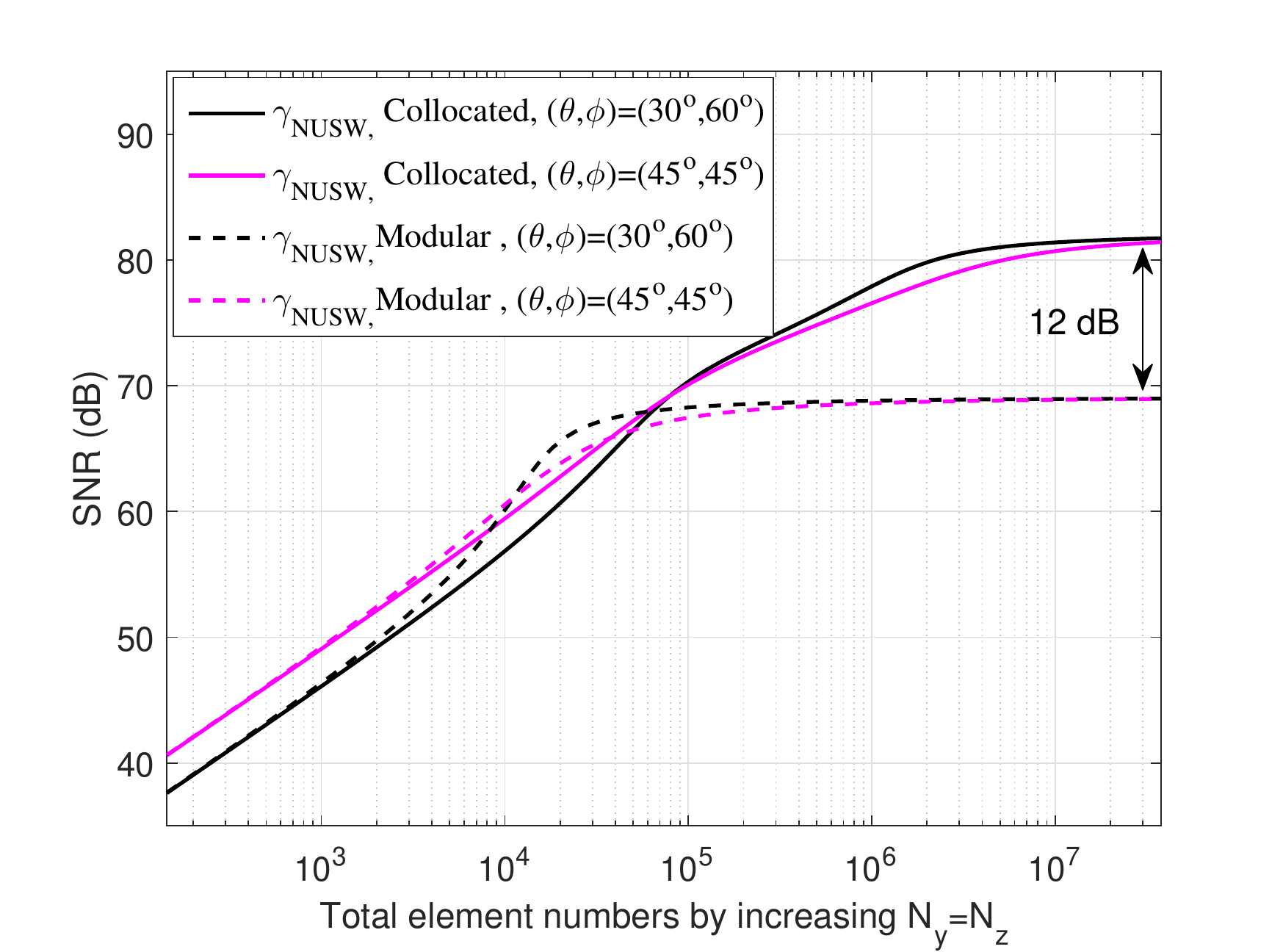}
\vspace{-0.3cm}
\caption{SNRs with different architectures versus the total number of array elements.} \label{picture15}
\end{centering}
\vspace{-0.3cm}
\end{figure}

\begin{figure}[t]
\begin{centering}
\includegraphics[scale=0.53]{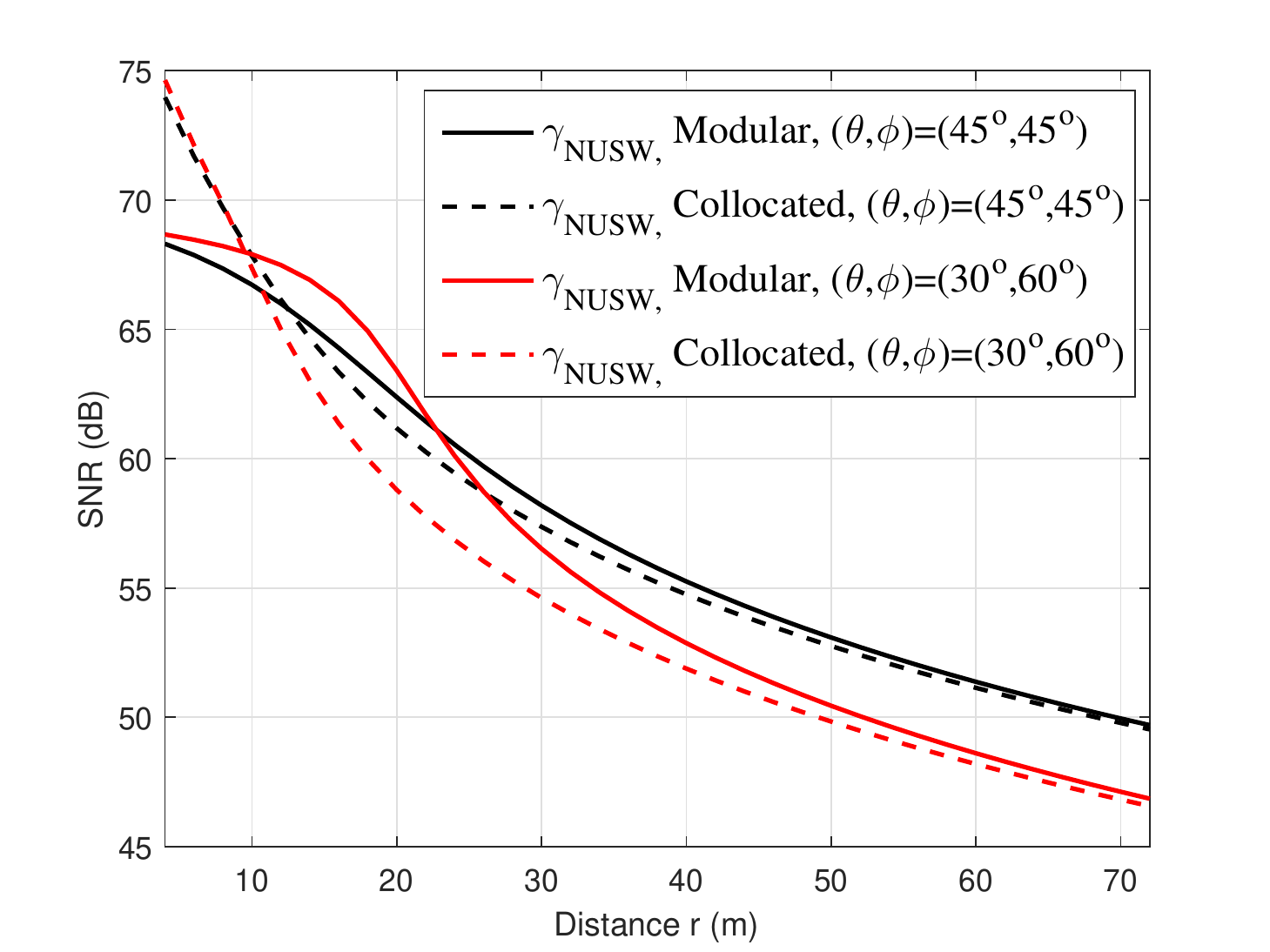}
\vspace{-0.3cm}
\caption{SNRs for different models versus the distance $r$.} \label{picture14}
\end{centering}
\vspace{-0.3cm}
\end{figure}

Fig.~\ref{picture13} plots the SNRs $\gamma_{\rm NUSW}$ and $\gamma_{\rm UPW}$
versus the module separation $D_y=D_z$ with two different zenith angles $\theta=30^\circ$ or $\theta=120^\circ$, where the azimuth angle is fixed to $\phi=30^\circ$. The module separation $D_y=D_z$ starting from $\frac{\lambda}{2}$ corresponds to the conventional collocated array architecture.
As shown from Fig.~\ref{picture13}, the SNR based on far-field UPW model is independent of $D_y$ and $D_z$.
By contrast, when $D_y$ and $D_z$ are relatively small, the SNRs based on NUSW model for different values of $\theta$ show different trends, while as $D_y$ and $D_z$ increase, the SNRs based on NUSW model for different values of $\theta$ exhibit the general decreasing trends. This illustrates that inter-module separation is an important parameter in the NUSW-based model, which greatly impacts on the system performance.\par

Fig.~\ref{picture15} plots the SNRs for different architectures versus the total number of array elements by increasing $N_y=N_z$ with two
different user
directions $(\theta,\phi)=(30^\circ,60^\circ)$ and $(\theta,\phi)=(45^\circ,45^\circ)$.
It is observed that for the two given user directions, when the physical size of the array is relatively small, the modular and collocated
architectures achieve similar SNR performance, while as the array size becomes large, the modular architecture provides slightly better performance than the collocated architecture initially while the opposite result is true when the array size becomes extremely large.
It is worth pointing out that the constant SNR difference between collocated architecture and modular architecture is about $12$ dB, which matches well with the theoretical result in (16).\par
As a further illustration, Fig.~\ref{picture14} plots the SNRs for different architectures versus the link distance $r$ with two user
directions $(\theta,\phi)=(30^\circ,60^\circ)$ and $(\theta,\phi)=(45^\circ,45^\circ)$. It is observed that the SNRs for all models generally decrease as the link distance $r$ increases, as expected. It is also observed from Fig.~\ref{picture14} that for the two given user directions, when $r$ is relatively small, the modular architecture has worse SNR performance than the collocated architecture, while as $r$ further increases, the modular architecture outperforms first and then the two models reach similar performance. \par

\begin{figure}[t]
\begin{centering}
\includegraphics[scale=0.54]{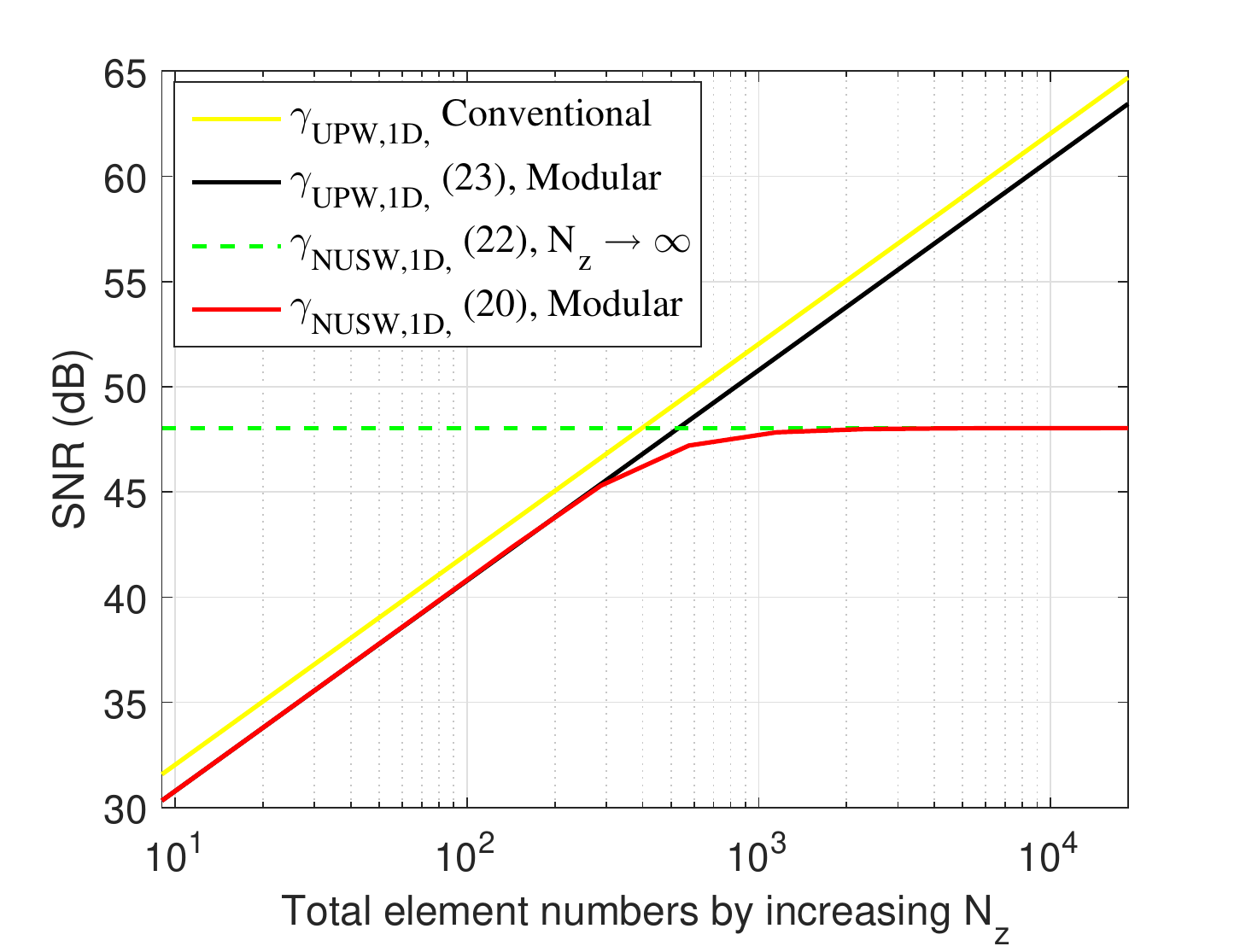}
\vspace{-0.3cm}
\caption{SNRs for XL-ULA communication versus the total number of array elements under  different models.} \label{picture16}
\end{centering}
\vspace{-0.3cm}
\end{figure}
Last, Fig.~\ref{picture16} shows the SNRs $\gamma_{\rm NUSW,1D}$ and $\gamma_{\rm UPW,1D}$ versus the total number of array elements $N_yN_zM$ with $N_y=1$ and $(\theta,\phi)=(60^\circ,30^\circ)$.
It is observed from Fig.~\ref{picture16}
that for relatively small array element numbers of $N_zM$,
both $\gamma_{\rm NUSW,1D}$ and $\gamma_{\rm UPW,1D}$ increase linearly with $N_zM$, which is consistent with the result in (23).
However, as $N_zM$ further increases with $N_z$, different from the result based on UPW modelling that grows linearly and eventually becomes unbounded,
the SNR result under the NUSW model approaches to a constant value.
It is also observed that the conventional $\gamma_{\rm UPW,1D}=\frac{\overline{P}N_z M\beta_0}{r^2}$  by ignoring variations of projected aperture across array elements over-estimates the result predicted by the proposed model.\par

\section{Conclusions}\label{g} \vspace{-1pt}
This paper studied the mathematical modelling and conducted the performance analysis for wireless communications with the modular XL-array, where different array modules are spaced with relatively large separations so as to keep conformal with practical building facades.
By taking into account the variations of signal phase, amplitude and projected aperture across array elements,
a generic array model is proposed. With the optimal MRC beamforming at the receiver,
the closed-form SNR expression was derived in terms of system geometric parameters, including the total planar array size, module separations, and the user's 3D location.
Besides, the asymptotic SNR scaling laws were
obtained as the number of modules along different dimensions goes to infinity. Furthermore,
our developed modelling and performance analysis were shown to include
the existing SNR results for the collocated XL-array, far-field UPW assumption and 1D modular XL-ULA architecture as special cases.
Extensive numerical results were provided to validate the importance of near-field NUSW modelling for accurately characterizing the performance of modular XL-array communications, which was shown to outperform the collocated XL-array for wider coverage due to more flexible array deployment.

\appendices

\section{Proof of Theorem 1}
Similar to the derivation in \cite{Zeng2021}, by observing the structure of triple summation in (8), we define the function
$f(a,b,c)\triangleq\frac{1}{[1-2a\Omega-2Kb\Omega-2K_yc\Phi+a^2+K^2b^2+2Kab+K_y^2c^2]^{\frac{3}{2}}}$,
which is a continuous function over the domain $\mathcal{A}=\left\{(a, b, c) \mid
-\frac{M \epsilon}{2} \leq a \leq \frac{M \epsilon}{2}, -\frac{N_{z} \epsilon}{2} \leq b \leq \frac{N_{z} \epsilon}{2},-\frac{N_{y} \epsilon}{2} \leq c \leq \frac{N_{y} \epsilon}{2}\right\}$.
Since $\epsilon \ll 1$, we have $f(a,b,c)\approx f(m\epsilon, n_z\epsilon, n_y\epsilon)$, $\forall (a,b,c) \in [(m-\frac{1}{2})\epsilon, (m+\frac{1}{2})\epsilon]
\times [(n_z-\frac{1}{2})\epsilon, (n_z+\frac{1}{2})\epsilon]
\times [(n_y-\frac{1}{2})\epsilon, (n_y+\frac{1}{2})\epsilon]$. Based on the concept of integral, we have
\begin{equation}\label{EQU-34} \vspace{-3pt}
\begin{split}
&\sum_{n_y=-\frac{N_y-1}{2}}^{\frac{N_y-1}{2}}\sum_{n_z=-\frac{N_z-1}{2}}^{\frac{N_z-1}{2}}
\sum_{m=-\frac{M-1}{2}}^{\frac{M-1}{2}}f(m\epsilon, n_z\epsilon, n_y\epsilon)\epsilon^3
\approx\iiint_{\mathcal{A}} f(a,b,c) \,da\,db\,dc.\\
\end{split}
\end{equation}\par
By substituting $f(a,b,c)$ into \eqref{EQU-34},
we have \eqref{EQU-35}, shown at the next page, where (a) and (b) follow from the integral formulas 2.264.5 and
2.02.10 in \cite{Gradshteyn2007}, respectively, as well as (c) holds owing to the integral formulas 2.01.18 and 2.284 in \cite{Gradshteyn2007}. By substituting \eqref{EQU-35} into \eqref{EQU-34}, the SNR result in \eqref{EQU-9} is obtained. The proof of Theorem 1 is thus completed.\par
\begin{footnotesize}
\begin{equation}\label{EQU-35} \vspace{-3pt}
\begin{split}
&\sum_{n_{y}=-\frac{N_{y}-1}{2}}^{\frac{N_{y}-1}{2}}
\sum_{n_{z}=-\frac{N_{z}-1}{2}}^{\frac{N_{z}-1}{2}}
\sum_{m=-\frac{M-1}{2}}^{\frac{M-1}{2}} \frac{1}{\left[1-2 m\epsilon\Omega-2Kn_z\epsilon\Omega-2 K_yn_y\epsilon\Phi+(m^2+K^2n_z^2+2Kmn_z+K_y^2n_y^2)\epsilon^2\right]^{\frac{3}{2}}}\\
&\approx \frac{1}{\epsilon^{3}} \int_{-\frac{N_y\epsilon}{2}}^{\frac{N_y\epsilon}{2}}
\int_{-\frac{N_z\epsilon}{2}}^{\frac{N_z\epsilon}{2}}\int_{-\frac{M\epsilon}{2}}^{\frac{M\epsilon}{2}}
\frac{1}{\left[1-2a\Omega-2Kb\Omega-2K_yc\Phi+a^2+K^2b^2+2Kab+K_y^2c^2\right]^{\frac{3}{2}}} d a d b d c\\
&\stackrel{(a)}{=} \frac{1}{\epsilon^{3}}\int_{-\frac{N_y\epsilon}{2}}^{\frac{N_y\epsilon}{2}}
\frac{1}{1-\Omega^2-2K_yc\Phi +K_y^2c^2}
\int_{-\frac{N_z\epsilon}{2}}^{\frac{N_z\epsilon}{2}}
\frac{Kb+\frac{M\epsilon}{2}-\Omega}
{\sqrt{1+\frac{1}{4}M^2\epsilon^2-M\epsilon\Omega -2K_yc\Phi +K_y^2c^2+(KM\epsilon-2K\Omega)b+K^2b^2}}\\
&\qquad\qquad \qquad \qquad \qquad \qquad \qquad \qquad-\frac{Kb-\frac{M\epsilon}{2}-\Omega}
{\sqrt{1+\frac{1}{4}M^2\epsilon^2+M\epsilon\Omega -2K_yc\Phi +K_y^2c^2-(KM\epsilon+2K\Omega)b+K^2b^2}} d b d c\\
&\stackrel{(b)}{=} \frac{1}{\epsilon^{3}K}\int_{-\frac{N_y\epsilon}{2}}^{\frac{N_y\epsilon}{2}}
\frac{1}{1-\Omega^2-2 K_yc\Phi +K_y^2c^2}
\left[\sqrt{\Psi^2+\left(\frac{K N_z\epsilon+M\epsilon}{2}-\Omega\right)^2+(K_yc-\Phi)^2}\right.\\
&\left.\qquad\qquad \qquad \qquad-\sqrt{\Psi^2+\left(\frac{K N_z\epsilon-M\epsilon}{2} -\Omega\right)^2+(K_yc-\Phi)^2}\right.
+ \sqrt{\Psi^2+\left(\frac{K N_z\epsilon+M\epsilon}{2}+\Omega\right)^2+(K_yc-\Phi)^2}\\
&\qquad\qquad \qquad \qquad \left.-\sqrt{\Psi^2+\left(\frac{K N_z\epsilon-M\epsilon}{2}+\Omega\right)^2+(K_yc-\Phi)^2}\right] d c\\
&\stackrel{(c)}{=}\frac{1}{\epsilon^{3}KK_y}\left\{\operatorname{arcsinh}\left(\frac{\frac{K_y N_y\epsilon}{2} -\Phi}{\sqrt{\Psi^2+\left(\frac{K N_z\epsilon+M\epsilon}{2} -\Omega\right)^2}}\right)+\operatorname{arcsinh}\left(\frac{\frac{K_y N_y\epsilon}{2} +\Phi}{\sqrt{\Psi^2+\left(\frac{K N_z\epsilon+M\epsilon}{2}-\Omega\right)^2}}\right)
+\frac{\frac{K N_z\epsilon+M\epsilon}{2}-\Omega}{\Psi}\times\right.\\
&\left[\arctan\left(\frac{(\frac{K_y N_y\epsilon}{2} -\Phi)(\frac{K N_z\epsilon+M\epsilon}{2} -\Omega)}
{\Psi\sqrt{\Psi^2+\left(\frac{K_y N_y\epsilon}{2} -\Phi\right)^2+\left(\frac{K N_z\epsilon+M\epsilon}{2} -\Omega \right)^2}} \right)
+\arctan\left(\frac{\left(\frac{K_y N_y\epsilon}{2} +\Phi\right)\left(\frac{K N_z\epsilon+M\epsilon}{2}-\Omega\right)}
{\Psi\sqrt{\Psi^2+\left(\frac{K_y N_y\epsilon}{2} +\Phi\right)^2+\left(\frac{K N_z\epsilon+M\epsilon}{2} -\Omega \right)^2}} \right)
\right]\\
&-\operatorname{arcsinh}\left(\frac{\frac{K_y N_y\epsilon}{2} -\Phi}{\sqrt{\Psi^2+\left(\frac{K N_z\epsilon-M\epsilon}{2}+\Omega\right)^2}}\right)
-\operatorname{arcsinh}\left(\frac{\frac{K_y N_y\epsilon}{2} +\Phi}{\sqrt{\Psi^2+\left(\frac{K N_z\epsilon-M\epsilon}{2}+\Omega\right)^2}}\right)
-\frac{\frac{K N_z\epsilon-M\epsilon}{2}+\Omega}
{\Psi}\times\\
&\left[\arctan\left(\frac{(\frac{K_y N_y\epsilon}{2}-\Phi)(\frac{K N_z\epsilon-M\epsilon}{2}+\Omega)}
{\Psi\sqrt{\Psi^2+\left(\frac{K_y N_y\epsilon}{2} -\Phi\right)^2+\left(\frac{K N_z\epsilon-M\epsilon}{2}+\Omega \right)^2}} \right)
+\arctan\left(\frac{\left(\frac{K_y N_y\epsilon}{2} +\Phi\right)\left(\frac{K N_z\epsilon-M\epsilon}{2}+\Omega\right)}
{\Psi\sqrt{\Psi^2+\left(\frac{K_y N_y\epsilon}{2} +\Phi\right)^2+\left(\frac{K N_2\epsilon-M\epsilon}{2}+\Omega \right)^2}} \right)\right]\\
&-\operatorname{arcsinh}\left(\frac{\frac{K_y N_y\epsilon}{2} -\Phi}{\sqrt{\Psi^2+\left(\frac{K N_z\epsilon-M\epsilon}{2}-\Omega\right)^2}}\right)
-\operatorname{arcsinh}\left(\frac{\frac{K_y N_y\epsilon}{2} +\Phi}{\sqrt{\Psi^2+\left(\frac{K N_z\epsilon-M\epsilon}{2}-\Omega\right)^2}}\right)
-\frac{\frac{K N_z\epsilon-M\epsilon}{2}-\Omega}
{\Psi}\times\\
&\left[\arctan\left(\frac{\left(\frac{K_y N_y\epsilon}{2}-\Phi\right)\left(\frac{K N_z\epsilon-M\epsilon}{2} -\Omega\right)}
{\Psi\sqrt{\Psi^2+\left(\frac{K_y N_y\epsilon}{2} -\Phi\right)^2+\left(\frac{K N_z\epsilon-M\epsilon}{2} -\Omega \right)^2}} \right)
+\arctan\left(\frac{\left(\frac{K_y N_y\epsilon}{2}+\Phi\right)\left(\frac{K N_z\epsilon-M\epsilon}{2}-\Omega\right)}
{\Psi\sqrt{\Psi^2+\left(\frac{K_y N_y\epsilon}{2} +\Phi\right)^2+\left(\frac{K N_z\epsilon-M\epsilon}{2} -\Omega \right)^2}} \right)\right]\\
&+\operatorname{arcsinh}\left(\frac{\frac{K_y N_y\epsilon}{2} -\Phi}{\sqrt{\Psi^2+\left(\frac{K N_z\epsilon+M\epsilon}{2}+\Omega\right)^2}}\right)
+\operatorname{arcsinh}\left(\frac{\frac{K_y N_y\epsilon}{2} +\Phi}{\sqrt{\Psi^2+\left(\frac{K N_z\epsilon+M\epsilon}{2}+\Omega\right)^2}}\right)+\frac{\frac{K N_z\epsilon+M\epsilon}{2} +\Omega}
{\Psi}\times\\
&\left.\left[
\arctan\left(\frac{\left(\frac{K_y N_y\epsilon}{2} -\Phi\right)\left(\frac{K N_z\epsilon+M\epsilon}{2} +\Omega\right)}
{\Psi\sqrt{\Psi^2+\left(\frac{K_y N_y\epsilon}{2} -\Phi\right)^2+\left(\frac{K N_z\epsilon+M\epsilon}{2} +\Omega \right)^2}} \right)
+\arctan\left(\frac{\left(\frac{K_y N_y\epsilon}{2} +\Phi\right)\left(\frac{K N_z\epsilon+M\epsilon}{2} +\Omega\right)}
{\Psi\sqrt{\Psi^2+\left(\frac{K_y N_y\epsilon}{2} +\Phi\right)^2+\left(\frac{K N_z\epsilon+M\epsilon}{2} +\Omega \right)^2}} \right)\right]\right\}.\\
\end{split}
\end{equation}
\end{footnotesize}
\section{Proof of Corollary 1}
Due to $D_y=D_z=d$, we have $K_y=1$ and $K=M$. Thus, the first term inside the bracket of (9) is
\begin{equation}\label{EQU-37} \vspace{-3pt}
\begin{split}
&F\left(\frac{\tilde{L}_y}{2r} -\Phi, \frac{\tilde{L}_z}{2r}-\Omega\right)=
F\left(\frac{N_yd}{2r} -\Phi, \frac{(MN_z+M)d}{2r}-\Omega\right) \\
&\stackrel{(a)}{\approx} F\left(\frac{N_yd}{2r} -\Phi, y\right)\Bigg|_{y=y_0}+\frac{Md}{2r}\frac{\partial F\left(\frac{N_yd}{2r} -\Phi, y\right)}{\partial y}\Bigg|_{y=y_0}\\
&=F\left(\frac{N_yd}{2r} -\Phi, \frac{MN_zd}{2r}-\Omega\right)
+\frac{Md}{2r\Psi}\arctan\left(\frac{(\frac{N_yd}{2r}-\Phi)(\frac{M N_zd}{2r}-\Omega)}
{\Psi\sqrt{\Psi^2+\left(\frac{N_yd}{2r} -\Phi\right)^2+\left(\frac{M N_zd}{2r} -\Omega \right)^2}} \right),\\
\end{split}
\end{equation}
where (a) follows from the first-order Taylor series expansion for
$y_0=\frac{MN_zd}{2r}-\Omega$ with relatively small $\frac{Md}{2r}$.
Similarly, we can obtain the approximations for other seven terms inside the bracket of \eqref{EQU-9}. By substituting all approximated results into \eqref{EQU-9},
we have \eqref{EQU-18}.
This thus completes the proof of Corollary 1.

\section{Proof of Corollary 2}
1) When $N_z\to \infty$, \eqref{EQU-9} can be expressed as the form of
the sum of four limit terms, i.e.,
\begin{equation}\label{EQU-45} \vspace{-3pt}
\begin{split}
\lim_{N_z\to\infty}\gamma_{\rm NUSW}
&=\frac{e\xi \bar{P} dr\Psi}{4\pi D_y[D_z+(M-1)d]}\left\{\lim_{N_z\to\infty}\left[F\left(B_1,\frac{Kd}{2r}N_z+B_3\right)
-F\left(B_1,\frac{Kd}{2r}N_z-B_4\right)\right]\right.\\
&+\left.\lim_{N_z\to\infty}\left[F\left(B_1,\frac{Kd}{2r}N_z+B_4
\right)-F\left(B_1,\frac{Kd}{2r}N_z-B_3 \right)\right]\right.\\
&+\lim_{N_z\to\infty}\left[F\left(B_2,\frac{Kd}{2r}N_z+B_3
\right)-F\left(B_2,\frac{Kd}{2r}N_z-B_4 \right)\right]\\
&+\left.\lim_{N_z\to\infty}\left[F\left(B_2,\frac{Kd}{2r}N_z+B_4
\right)-F\left(B_2,\frac{Kd}{2r}N_z-B_3 \right)\right]
 \right\},\\
\end{split}
\end{equation}
where $B_1=\frac{\tilde{L}_y}{2r}-\Phi$, $B_2=\frac{\tilde{L}_y}{2r}+\Phi$, $B_3=\frac{\tilde{L}_e}{2r}-\Omega$ and $B_4=\frac{\tilde{L}_e}{2r}+\Omega$. For the first limit term inside the bracket, let $g_1(N_z)=\frac{Kd}{2r}N_z+B_3$,
$g_2(N_z)=\frac{Kd}{2r}N_z-B_4$, $g_3(N_z)=\frac{Kd}{2r}N_z-\Omega$, and $g_4(N_z)=\sqrt{\Psi^2+B_1^2+g_3^2(N_z)}$ whose highest order is one, and thus
we have
\begin{equation}\label{EQU-46} \vspace{-3pt}
\begin{split}
&\lim_{N_z \to \infty}\left[F\left(B_1,g_1(N_z)\right)-\right.
\left.F\left(B_1,g_2(N_z)\right)\right]\\
&=\lim_{N_z \to \infty}
\left[\operatorname{arcsinh}\left(\frac{B_1}{\sqrt{\Psi^2+g_1^2(N_z)}}\right)-\operatorname{arcsinh}\left(\frac{B_1}{\sqrt{\Psi^2+g^2_2(N_z)}}\right)\right]\\
&+\lim_{N_z \to \infty}\frac{g_2(N_z)}{\Psi}
\left[
\arctan\left(\frac{B_1g_1(N_z)}
{\Psi\sqrt{\Psi^2+B_1^2+g_1^2(N_z)}} \right)-\arctan\left(\frac{B_1g_2(N_z)}
{\Psi\sqrt{\Psi^2+B_1^2+g_2^2(N_z)}} \right)\right]\\
&+\lim_{N_z \to \infty}\frac{\tilde{L}_e}{r\Psi}
\arctan\left(\frac{B_1g_1(N_z)}
{\Psi\sqrt{\Psi^2+B_1^2+g_1^2(N_z)}} \right)\\
&\stackrel{(a)}{=} \frac{\tilde{L}_e}{r}\left[\lim_{N_z \to \infty}\frac{-B_1g_3(N_z)}{\left[\Psi^2+g_3^2(N_z)\right]
g_4(N_z)}
+\lim_{N_z \to \infty}\frac{\left[\Psi^2B_1+B_1^3\right]g_2(N_z)}
{\left[B_1^2g_3^2(N_z)+\Psi^2
g_4^2(N_z)\right]
g_4(N_z)}\right]+\frac{\tilde{L}_e}{r\Psi}
\arctan\left(\frac{B_1}{\Psi} \right)\\
&\stackrel{(b)}{=}\frac{\tilde{L}_e}{r\Psi}
\arctan\left(\frac{B_1}{\Psi} \right),
\end{split}
\end{equation}
where (a) holds according to the first-order Taylor series expansion with relatively small $\frac{Md}{2r}$, as well as (b) follows from fractional polynomial limit theorem. Due to the similar form of four terms inside the bracket, we have \eqref{EQU-19}.\par
2) When $N_y\to \infty$, \eqref{EQU-9} can be also expressed as the form of the sum of four limit terms, i.e.,
\begin{equation}\label{EQU-45} \vspace{-3pt}
\begin{split}
\lim_{N_y\to\infty}\gamma_{\rm NUSW}
&\approx\frac{e\xi \bar{P} dr\Psi}{4\pi D_y[D_z+(M-1)d]}\left\{\lim_{N_y\to\infty}\left[F\left(\frac{K_yd}{2r}N_y-\Phi,B_5\right)-
F\left(\frac{K_yd}{2r}N_y-\Phi,B_6\right)\right]\right.\\
&+\lim_{N_y\to\infty}\left[F\left(\frac{K_yd}{2r}N_y-\Phi,B_7
\right)-F\left(\frac{K_yd}{2r}N_y-\Phi,B_8\right)\right]\\
&+\lim_{N_y\to\infty}\left[F\left(\frac{K_yd}{2r}N_y+\Phi,B_5
\right)-F\left(\frac{K_yd}{2r}N_y+\Phi,B_6\right)\right]\\
&+\lim_{N_y\to\infty}\left[F\left(\frac{K_yd}{2r}N_y+\Phi,B_7
\right)-
\left.F\left(\frac{K_yd}{2r}N_y+\Phi,B_8\right)\right]
 \right\},\\
\end{split}
\end{equation}
where $B_5=\frac{\tilde{L}_z}{2r}-\Omega$, $B_6=\frac{\hat{L}_z}{2r}-\Omega$,
$B_7=\frac{\tilde{L}_z}{2r}+\Omega$ and $B_8=\frac{\hat{L}_z}{2r}+\Omega$. For the first limit term inside the bracket, let $h_1(N_y)=\frac{K_yd}{2r}N_y-\Phi$, $h_2(N_y)=\sqrt{\Psi^2+B_9^2+h_1^2(N_y)}$ and $B_9=\frac{\tilde{L}_z-\tilde{L}_e}{2r}-\Omega$, we have
\begin{equation}\label{EQU-461} \vspace{-3pt}
\begin{split}
&\lim_{N_y \to \infty}\left[F\left(h_1(N_y), B_5\right)-F\left(h_1(N_y), B_6\right)\right]\\
&=\lim_{N_y \to \infty}
\left[\operatorname{arcsinh}\left(\frac{h_1(N_y)}{\sqrt{\Psi^2+B_5^2}}\right)\right.
\left.-\operatorname{arcsinh}\left(\frac{h_1(N_y)}{\sqrt{\Psi^2+B_6^2}}\right)\right]\\
&+\lim_{N_y \to \infty}\frac{B_6}{\Psi}
\left[
\arctan\left(\frac{B_5h_1(N_y)}
{\Psi\sqrt{\Psi^2+B_5^2+h_1^2(N_y)}} \right)-\arctan\left(\frac{B_6h_1(N_y)}
{\Psi\sqrt{\Psi^2+B_6^2+h_1^2(N_y)}} \right)\right]\\
&+\lim_{N_y \to \infty}\frac{\tilde{L}_e}{r\Psi}
\arctan\left(\frac{B_5 h_1(N_y)}
{\Psi\sqrt{\Psi^2+B_5^2+h_1^2(N_y)}} \right)\\
&\stackrel{(a)}{=} \frac{\tilde{L}_e}{r}\left[\lim_{N_y \to \infty}\frac{-B_9 h_1(N_y)}{\left(\Psi^2+B_9^2\right)
h_2(N_y)}\right.+\lim_{N_y \to \infty}
\left.
\frac{\left[\Psi^2h_1(N_y)+h_1^3(N_y)\right]B_6}
{\left[B_9^2 h_1^2(N_y)+\Psi^2
h_2^2(N_y)\right]
h_2(N_y)}\right]+\frac{\tilde{L}_e}{r\Psi}
\arctan\left(\frac{B_5}{\Psi} \right)\\
&\stackrel{(b)}{=}-\frac{\tilde{L}_e^2}{2r^2(\Psi^2+B_9^2)}+\frac{\tilde{L}_e}{r\Psi}
\arctan\left(\frac{B_5}{\Psi} \right),
\end{split}
\end{equation}
where (a) and (b) follow from conditions (a) and (b) of \eqref{EQU-46}, respectively. Due to the same form of four terms inside the bracket, we have \eqref{EQU-1911}.

3) When $N_y, N_z\to \infty$ with $N_z=\tau N_y$ and $0<\tau<\infty$, \eqref{EQU-9} can be formulated as the sum form of four limit terms, i.e.,
\begin{equation}\label{EQU-45} \vspace{-3pt}
\begin{split}
&\lim_{N_y, N_z\to\infty}\gamma_{\rm NUSW}
\approx\frac{e\xi \bar{P} dr\Psi}{4\pi D_y[D_z+(M-1)d]}\times\\
&\left\{\lim_{N_y\to\infty}\left[F\left(\frac{K_yd}{2r}N_y -\Phi, \frac{Kd \tau}{2r}N_y+B_3\right)\right. -F\left(\frac{K_yd}{2r}N_y-\Phi, \frac{Kd \tau}{2r}N_y-B_4\right)\right]\\
&+\left\{\lim_{N_y\to\infty}\left[F\left(\frac{K_yd}{2r}N_y-\Phi, \frac{Kd \tau}{2r}N_y+B_4
\right)-F\left(\frac{K_yd}{2r}N_y-\Phi,
\frac{Kd \tau}{2r}N_y-B_3 \right)\right]\right\}\\
&+\left\{\lim_{N_y\to\infty}\left[F\left(\frac{K_yd}{2r}N_y+\Phi, \frac{Kd \tau}{2r}N_y+B_3
\right)\right.-F\left(\frac{K_yd}{2r}N_y+\Phi,
\frac{Kd \tau}{2r}N_y-B_4 \right)\right]\\
&+\left\{\lim_{N_y\to\infty}\left[F\left(\frac{K_yd}{2r}N_y+\Phi, \frac{Kd \tau}{2r}N_y+B_4
\right)-F\left(\frac{K_yd}{2r}N_y+\Phi,
\frac{Kd \tau}{2r}N_y-B_3 \right)\right]
 \right\}.\\
\end{split}
\end{equation}\par
For the first limit term inside the bracket, by assuming $h_3(N_y)=\sqrt{\Psi^2+h^2_1(N_y)+g_3^2(\tau N_y)}$, we have
\begin{equation}\label{EQU-462} \vspace{-3pt}
\begin{split}
&\lim_{N_y \to \infty}\left[F\left(h_1(N_y), g_1(\tau N_y)\right)-F\left(h_1(N_y), g_2(\tau N_y)\right)\right]\\
&=\lim_{N_y \to \infty}
\left[\operatorname{arcsinh}\left(\frac{h_1(N_y)}{\sqrt{\Psi^2+g_1^2(\tau N_y)}}\right)\right.\left.-\operatorname{arcsinh}\left(\frac{h_1(N_y)}{\sqrt{\Psi^2+g_2^2(\tau N_y)}}\right)\right]\\
&+\lim_{N_y \to \infty}\frac{g_2(\tau N_y)}{\Psi}
\left[
\arctan\left(\frac{h_1(N_y)g_1(\tau N_y)}
{\Psi\sqrt{\Psi^2+h^2_1(N_y)+g_1^2(\tau N_y)}} \right)\right.\\
&\left.-\arctan\left(\frac{h_1(N_y)g_2(\tau N_y)}
{\Psi\sqrt{\Psi^2+h^2_1(N_y)+g_2^2(\tau N_y)}} \right)\right]+\lim_{N_y \to \infty}\frac{\tilde{L}_e}{r\Psi}
\arctan\left(\frac{h_1(N_y)g_1(\tau N_y)}
{\Psi\sqrt{\Psi^2+h_1(N_y)^2+g^2_1(\tau N_y)}} \right)\\
&\stackrel{(a)}{=} \frac{\tilde{L}_e}{r}\left[\lim_{N_y \to \infty}\frac{-h_1(N_y)g_3(\tau N_y)}{\left[\Psi^2+g^2_3(\tau N_y)\right]
h_3(N_y)}
+\lim_{N_y \to \infty}\frac{\left[\Psi^2h_1(N_y)+h^3_1(N_y)\right]g_2(\tau N_y)}
{\left[h_1(N_y)^2g^2_3(\tau N_y)+\Psi^2
h_3^2(N_y)\right]
h_3(N_y)}\right]+\frac{\pi \tilde{L}_e}{2 r\Psi}\\
&\stackrel{(b)}{=}\frac{\pi \tilde{L}_e}{2r\Psi},
\end{split}
\end{equation}
where (a) holds thanks to the first-order Taylor series expansion with relatively small $\frac{\tilde{L}_e}{2r}$ and $\lim _{x, y \rightarrow \infty} \arctan \left(\frac{x y}{\Psi \sqrt{\Psi+x^{2}+y^{2}}}\right)=\frac{\pi}{2}$, and (b) follows from fractional polynomial limit theorem as well.
Since other three terms inside the bracket has the similar form, we have \eqref{EQU-192}.
The proof of Corollary 2 is therefore completed.

\section{Proof of Corollary 3}
When $r\Psi \gg \tilde{L}_y$ and $r\Psi \gg \tilde{L}_z$, we have $\frac{\tilde{L}_y}{2r\Psi}\ll 1$ and $\frac{\tilde{L}_e}{2r\Psi}\le\frac{\tilde{L}_z}{2r\Psi}\ll 1$. \eqref{EQU-9} is thus expressed as
\begin{equation}\label{EQU-38} \vspace{-3pt}
\begin{split}
&\gamma_{\rm NUSW}\approx\frac{e\xi\bar{P}d r\Psi }{4 \pi D_y[D_z+(M-1)d]}
\left[F_2\left(\frac{\tilde{L}_y}{2r\Psi} -\frac{\Phi}{\Psi}, \frac{\tilde{L}_z}{2r\Psi}-\frac{\Omega}{\Psi}\right)\right.
-F_2\left(\frac{\tilde{L}_y}{2r\Psi}-\frac{\Phi}{\Psi}, \frac{\hat{L}_z}{2r\Psi}-\frac{\Omega}{\Psi}\right)\\
&+F_2\left(\frac{\tilde{L}_y}{2r\Psi}-\frac{\Phi}{\Psi}, \frac{\tilde{L}_z}{2r\Psi}+\frac{\Omega}{\Psi}
\right)
-F_2\left(\frac{\tilde{L}_y}{2r\Psi}-\frac{\Phi}{\Psi},\frac{\hat{L}_z}{2r\Psi}+\frac{\Omega}{\Psi} \right)
+F_2\left(\frac{\tilde{L}_y}{2r\Psi}+\frac{\Phi}{\Psi}, \frac{\tilde{L}_z}{2r\Psi}-\frac{\Omega}{\Psi}
\right)\\
&-F_2\left(\frac{\tilde{L}_y}{2r\Psi}+\frac{\Phi}{\Psi},
\frac{\hat{L}_z}{2r\Psi}-\frac{\Omega}{\Psi} \right)+F_2\left(\frac{\tilde{L}_y}{2r\Psi}+\frac{\Phi}{\Psi}, \frac{\tilde{L}_z}{2r\Psi}+\frac{\Omega}{\Psi}
\right)
\left.-F_2\left(\frac{\tilde{L}_y}{2r\Psi}+\frac{\Phi}{\Psi},
\frac{\hat{L}_z}{2r\Psi}+\frac{\Omega}{\Psi} \right)
 \right],\\
\end{split}
\end{equation}
where $F_2(x,y)\triangleq \operatorname{arcsinh}\left(\frac{x}{\sqrt{1+y^2}}\right)
+y\arctan\left(\frac{xy}{\sqrt{1+x^2+y^2}} \right)$.\par
Similar to the proof of Corollary 2,
by using the first-order Taylor series expansion of
$F_2\left(\frac{\tilde{L}_y}{2r\Psi} -\frac{\Phi}{\Psi}, \frac{\tilde{L}_z}{2r\Psi}-\frac{\Omega}{\Psi}\right)$ for
$y_0=\frac{\tilde{L}_z-\tilde{L}_e}{2r\Psi}-\frac{\Omega}{\Psi}$ with $\frac{\tilde{L}_e}{2r\Psi}\ll 1$, we have
\begin{equation}\label{EQU-39} \vspace{-3pt}
\begin{split}
&F_2\left(\frac{\tilde{L}_y}{2r\Psi} -\frac{\Phi}{\Psi}, \frac{\tilde{L}_z}{2r\Psi}-\frac{\Omega}{\Psi}\right)
\approx F_2\left(\frac{\tilde{L}_y}{2r\Psi} -\frac{\Phi}{\Psi}, y\right)\Bigg|_{y=y_0}+\frac{\tilde{L}_e}{2r\Psi}\frac{\partial F_2\left(\frac{\tilde{L}_y}{2r\Psi} -\frac{\Phi}{\Psi}, y\right)}{\partial y}\Bigg|_{y=y_0}\\
&=F_2\left(\frac{\tilde{L}_y}{2r\Psi} -\frac{\Phi}{\Psi}, \frac{\tilde{L}_z-\tilde{L}_e}{2r\Psi}-\frac{\Omega}{\Psi}\right)
+\frac{\tilde{L}_e}{2r\Psi}\arctan\left(\frac{\left(\frac{\tilde{L}_y}{2r\Psi}-\frac{\Phi}{\Psi}\right)\left(\frac{\tilde{L}_z-\tilde{L}_e}{2r\Psi}-\frac{\Omega}{\Psi}\right)}
{\sqrt{1+\left(\frac{\tilde{L}_y}{2r\Psi} -\frac{\Phi}{\Psi}\right)^2+\left(\frac{\tilde{L}_z-\tilde{L}_e}{2r\Psi} -\frac{\Omega}{\Psi}\right)^2}} \right)\\
&\stackrel{(a)}{\approx} F_2\left(\frac{\tilde{L}_y}{2r\Psi} -\frac{\Phi}{\Psi}, \frac{\tilde{L}_z-\tilde{L}_e}{2r\Psi}-\frac{\Omega}{\Psi}\right)
+\frac{\tilde{L}_e}{2r\Psi}\frac{\left(\frac{\tilde{L}_y}{2r\Psi}-\frac{\Phi}{\Psi}\right)\left(\frac{\tilde{L}_z-\tilde{L}_e}{2r\Psi}-\frac{\Omega}{\Psi}\right)}
{\sqrt{1+\left(\frac{\tilde{L}_y}{2r\Psi} -\frac{\Phi}{\Psi}\right)^2+\left(\frac{\tilde{L}_z-\tilde{L}_e}{2r\Psi} -\frac{\Omega}{\Psi}\right)^2}},\\
\end{split}
\end{equation}
where (a) holds due to $\arctan x \approx x$ for $|x|\ll1$ and the condition $\frac{(\frac{\tilde{L}_y}{2r\Psi}-\frac{\Phi}{\Psi})(\frac{\tilde{L}_z-\tilde{L}_e}{2r\Psi}-\frac{\Omega}{\Psi})}{\sqrt{1+\left(\frac{\tilde{L}_y}{2r\Psi} - \frac{\Phi}{\Psi}\right)^{2}+\left(\frac{\tilde{L}_y-\tilde{L}_e}{2r\Psi}-\frac{\Omega}{\Psi}\right)^{2}}} \ll 1$.
Similarly, we can obtain the approximations for other seven terms inside the bracket of \eqref{EQU-38}. Furthermore, by substituting all approximated results into \eqref{EQU-38}, we have
\begin{equation}\label{EQU-40} \vspace{-3pt}
\begin{split}
\gamma_{\rm NUSW}&\approx
\frac{e\xi \bar{P}M}{4 \pi KK_y}\left[F_3\left(\frac{\tilde{L}_y}{2 r\Psi}-\frac{\Phi}{\Psi}, \frac{\tilde{L}_z-\tilde{L}_e}{2 r\Psi}-\frac{\Omega}{\Psi}\right)\right.+F_3\left(\frac{\tilde{L}_y}{2 r\Psi}-\frac{\Phi}{\Psi}, \frac{\tilde{L}_z-\tilde{L}_e}{2 r\Psi}+\frac{\Omega}{\Psi}\right)\\
&+F_3\left(\frac{\tilde{L}_y}{2 r\Psi}+\frac{\Phi}{\Psi}, \frac{\tilde{L}_z-\tilde{L}_e}{2 r\Psi}-\frac{\Omega}{\Psi}\right)\left.+F_3\left(\frac{\tilde{L}_y}{2 r\Psi}+\frac{\Phi}{\Psi}, \frac{\tilde{L}_z-\tilde{L}_e}{2 r\Psi}+\frac{\Omega}{\Psi}\right)\right],
\end{split}
\end{equation}
where $F_3(x,y)=\frac{xy}{\sqrt{1+x^2+y^2}}$.\par

Then, we express the denominator
of the first term as a function of $u \triangleq \frac{\tilde{L}_y}{2 r \Psi}$ and $v \triangleq \frac{\tilde{L}_z-\tilde{L}_e}{2 r \Psi}$,
defined as $g(u,v)\triangleq \sqrt{1+(u-\frac{\Phi}{\Psi})^2+(v-\frac{\Omega}{\Psi})^2}$. By applying the first-order Taylor approximation for small $u$ and $v$, it follows that
\begin{equation}\label{EQU-41} \vspace{-3pt}
g(u,v)\approx C_1-\frac{\frac{\Phi}{\Psi}u}{C_1}-\frac{\frac{\Omega}{\Psi}v}{C_1}, C_1 \triangleq \sqrt{1+\frac{\Phi^2}{\Psi^2}+\frac{\Omega^2}{\Psi^2}}.
\end{equation}\par
Similar to other
three terms inside the bracket, \eqref{EQU-40} can be approximated as
\begin{equation}\label{EQU-42} \vspace{-3pt}
\begin{split}
&\gamma_{\rm NUSW}\approx
 \frac{e\xi \bar{P}M}{4 \pi KK_y}\left[F_4\left(-\frac{\Phi}{\Psi}, -\frac{\Omega}{\Psi}\right)+F_4\left(-\frac{\Phi}{\Psi}, \frac{\Omega}{\Psi}\right)+F_4\left(\frac{\Phi}{\Psi}, -\frac{\Omega}{\Psi}\right)+F_4\left(\frac{\Phi}{\Psi}, \frac{\Omega}{\Psi}\right)\right],
\end{split}
\end{equation}
where $F_4\triangleq \frac{\left(u+x\right)\left(v+y\right)}{C_1+\frac{x}{C_1} u+\frac{y}{C_1} v}$.
By summing $F_{4}\left(-\frac{\Phi}{\Psi},-\frac{\Omega}{\Psi}\right)$ and $F_{4}\left(\frac{\Phi}{\Psi},\frac{\Omega}{\Psi}\right)$ inside the bracket at first, we have
\begin{equation}\label{EQU-43} \vspace{-3pt}
\begin{split}
\begin{aligned}
F_4\left(-\frac{\Phi}{\Psi},-\frac{\Omega}{\Psi}\right)+F_4\left(\frac{\Phi}{\Psi}, \frac{\Omega}{\Psi}\right)
&=2 \frac{\left(C_1-\frac{\Omega^{2}}{C_1 \Psi^{2}}-\frac{\Phi^{2}}{C_1 \Psi^{2}}\right) uv+C_1 \frac{\Phi \Omega}{\Psi^{2}}-\frac{\Phi \Omega}{C_1 \Psi^{2}}\left(u^2+v^2\right)}{C_1^{2}-\left(\frac{\Phi}{C_1 \Psi} u+\frac{\Omega}{C_1 \Psi} v\right)^{2}} \\
&\stackrel{(a)}{\approx} 2 \frac{\left(C_1-\frac{\Omega^{2}}{C_1 \Psi^{2}}-\frac{\Phi^{2}}{C_1 \Psi^{2}}\right)uv+C_1 \frac{\Phi \Omega}{\Psi^{2}}-\frac{\Phi \Omega}{C_1 \Psi^{2}}\left(u^2+v^2\right)}{C_1^{2}},
\end{aligned}
\end{split}
\end{equation}
where (a) holds due to the fact $u\ll 1$ and $v\ll 1$.
Similarly, the sum of $F_{4}\left(\frac{\Phi}{\Psi},-\frac{\Omega}{\Psi}\right)$ and $F_{4}\left(-\frac{\Phi}{\Psi},\frac{\Omega}{\Psi}\right)$ can be
 obtained. After substituting all these results into \eqref{EQU-42}, we have
\begin{equation}\label{EQU-44} \vspace{-3pt}
\begin{split}
\begin{aligned}
\gamma_{\rm NUSW} & \approx \frac{e\xi \bar{P}M}{4 \pi K K_y} 4 \frac{\left(C_1-\frac{\Omega^{2}}{C_1 \Psi^{2}}-\frac{\Phi^{2}}{C_1 \Psi^{2}}\right)uv}{C_1^{2}}=\frac{\bar{P} N_yN_zMeA}{4 \pi r^{2} \Psi^{2}} \frac{\left(C_1^{2}-\frac{\Omega^{2}}{\Psi^{2}}-\frac{\Phi^{2}}{\Psi^{2}}\right)}{C_1^{3}} \\
&=\frac{\bar{P} N_yN_zMeA}{4 \pi r^{2} \Psi^{2}} \frac{1}{\left(1+\frac{\Omega^{2}}{\Psi^{2}}+\frac{\Phi^{2}}{\Psi^{2}}\right)^{\frac{3}{2}}}=\frac{\bar{P} N_yN_zMeA\Psi }{4 \pi r^{2}}.
\end{aligned}
\end{split}
\end{equation}\par
This thus completes the proof of Corollary 3.

\section{Proof of Corollary 4}
When $N_y=1$ and $K_y=1$, we express the first term inside the bracket of \eqref{EQU-9} as
\begin{equation}\label{EQU-48} \vspace{-3pt}
\begin{split}
&F\left(\frac{\tilde{L}_y}{2r} -\Phi, \frac{\tilde{L}_z}{2r}-\Omega\right)=F\left(\frac{d}{2r}-\Phi, \frac{\tilde{L}_z}{2r}-\Omega\right)\\
&\stackrel{(a)}{\approx}F\left(x, \frac{\tilde{L}_z}{2r}-\Omega\right)\bigg|_{x=x_0}+
\frac{d}{2r}\frac{\partial F\left(x,
\frac{\tilde{L}_z}{2r}-\Omega\right)}{\partial x}\bigg|_{x=x_0}\\
&=F\left(-\Phi, \frac{\tilde{L}_z}{2r}-\Omega\right)+
\frac{d}{2r\sin^2\theta}\sqrt{\sin^2\theta+
\left(\frac{\tilde{L}_z}{2r}-\Omega \right)^2},
\end{split}
\end{equation}
where (a) follows from the first-order Taylor series expansion for
$x_0=-\Phi$ with relatively small $\frac{d}{2r}$.
Similarly, we can obtain the approximations for other seven terms inside the bracket of \eqref{EQU-9}. By substituting all approximated results into \eqref{EQU-9}, we have \eqref{EQU-23}. Therefore, the proof of Corollary 4 is completed.

\end{document}